# Dopant-free molecular hole transport material that mediates a 20% power conversion efficiency in a perovskite solar cell

Yang Cao[1,4,†], Yunlong Li[2,†], Thomas Morrissey[1,4], Brian Lam[1], Brian O. Patrick[1], David J. Dvorak[4], Zhicheng Xia[1], Timothy L. Kelly[2,*], Curtis P. Berlinguette[1,3,4,*]

[1]Department of Chemistry, The University of British Columbia, 2036 Main Mall, Vancouver, British Columbia, V6T 1Z1, Canada.

[2]Department of Chemistry, University of Saskatchewan, 110 Science Place, Saskatoon, Saskatchewan, S7N 5C9, Canada.

[3]Department of Chemical and Biological Engineering, The University of British Columbia, 2360 East Mall, Vancouver, British Columbia, V6T 1Z3, Canada.

[4]Stewart Blusson Quantum Matter Institute, The University of British Columbia, 2355 East Mall, Vancouver, British Columbia, V6T 1Z4, Canada.

* Correspondence to tim.kelly@usask.ca and cberling@chem.ubc.ca

[†]These authors contributed equally to this work

**Abstract:**

Organic molecular hole-transport materials (HTMs) are appealing for the scalable manufacture of perovskite solar cells (PSCs) because they are easier to reproducibly prepare in high purity than polymeric and inorganic HTMs. There is also a need to construct PSCs without dopants and additives to avoid formidable engineering and stability issues. We report here a power conversion efficiency (PCE) of 20.6% with a molecular HTM in an inverted (p-i-n) PSC without any dopants or interlayers. This new benchmark was made possible by the discovery that annealing a spiro-based dopant-free HTM (denoted **DFH**) containing redox-active triphenyl amine (TPA) units undergoes preferential molecular organization normal to the substrate. This structural order, governed by the strong intermolecular interactions of the **DFH** dioxane groups, affords high intrinsic hole mobility ($1 \times 10^{-3}$ cm$^2 \cdot$V$^{-1} \cdot$s$^{-1}$). Annealing films of **DFH** also enables the growth of large perovskite grains (up to 2 µm) that minimize charge recombination in the PSC. **DFH** can also be isolated at a fraction of the cost of any other organic HTM.

Metal halide perovskite solar cells (PSCs) rely on hole-transport materials (HTMs) to efficiently extract holes from the perovskite layer and minimize charge recombination at the anode.[1–4] A wide range of inorganic metal oxides,[5] organic π-conjugated polymers[6] and organic small molecules[7] have proven to be effective HTMs in the PSC (Fig. 1). Organic small molecules are particularly appealing because they offer acute control of physicochemical properties, and are relatively easy to synthesize, purify, and process.[7–9] Moreover, the hole transport layer of the state-of-the-art PSC is based on organic small molecules.[10]

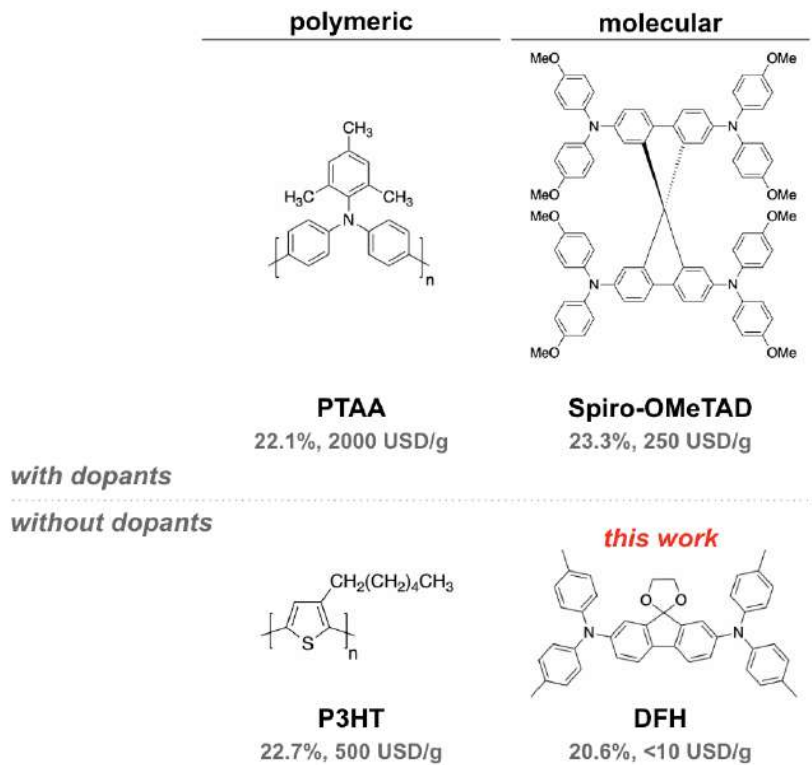

**Fig. 1.** Benchmark *PCE*s for devices containing organic HTMs with and without dopants. **DFH** can be synthesized at a fraction of the cost of the other HTMs and does not require interlayers (P3HT requires interlayers to reach 20%). The champion PSC device with an inorganic HTM is 20.6% (not listed).

A key challenge of molecular HTMs is that they usually require dopants to reach the high conductivities necessary for high device power conversion efficiencies (*PCE*s).[10–12] Dopants can compromise device performance by accelerating deleterious moisture permeation, ion migration, and interfacial charge recombination during device operation.[12–16] This situation can be addressed in part with barrier layers,[17,18] but a simpler solution is to design HTMs that do not require dopants. This has prompted the design of dopant-free molecular HTMs consisting of large planar π-stacked [19–22] or π-conjugated donor-acceptor molecules.[23–26] While dopant-free polymeric HTMs have yielded *PCE*s as high as 22.7% in devices containing additional interlayers,[27] these interlayers complicate the device fabrication process and can reduce the thermal stability of the cell.[10] Molecular HTMs have not previously reached the 20% *PCE* threshold without the use of dopants.[19,28,29]

We report herein a new benchmark for dopant-free molecular HTMs by showing that devices containing $N^2,N^2,N^7,N^7$-tetra-*p*-tolylspiro[fluorene-9,2'-[1,3]dioxolane]-2,7-diamine (**DFH**) as the HTM can yield a *PCE* of 20.6% (Figs. 1 and 2a). This breakthrough was realized in an inverted PSC architecture *without the assistance of p-dopants or interlayers*, and made possible by designing **DFH**: (i) with an appropriately positioned HOMO energy for efficient charge extraction from the perovskite layer; and (ii) with functional groups that encourage anisotropic molecular ordering of the film to mediate high electronic conductivity and hole mobility normal to the perovskite layer. This layer also accommodates the growth of large crystalline domains for the perovskite layer (Fig. 2B). Moreover, this rationally designed **DFH** can be isolated by the reaction of inexpensive reagents followed by a facile purification process that could enable it to be scaled at a cost of <$10/g (Tables S1 and S2). These economics, which cannot be matched by any champion organic HTM (Fig. 1), compare favorably to the inexpensive sol-gel chemistry used to make inorganic HTMs.[30,31] This feature is important because molecular organic HTMs are viewed to be generally easier to scale since they do not suffer from the same batch-to-batch variability as inorganic and polymeric HTMs.[32,33]

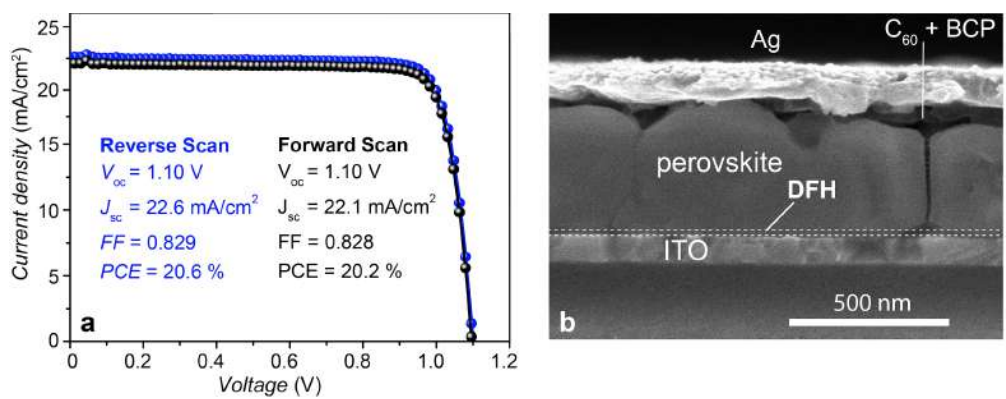

**Fig. 2.** (a) Current-voltage traces of the champion device in the forward and reverse directions (scan rate = 160 mV/s). (b) Cross-sectional scanning electron microscope images of an inverted (p-i-n) PSC containing **DFH** as the HTM (ITO = indium tin oxide; BCP = bathocuproine).

The structure of **DFH** features two fluorene-bridged triphenylamine (TPA) moieties linked to a 1,3-dioxolane group through a spiro carbon center (Fig. 1). The HTM is formed through a successive Buchwald–Hartwig amination of 2,7-dibromo-9-fluorenone and acid-catalyzed condensation with ethylene glycol. Products from each reaction can be purified by recrystallization rather than column chromatography to yield **DFH** in 73% overall reaction yield.[8] The design of **DFH** was inspired by the knowledge that spiro-type HTMs exhibit low redox reorganization energies, high glass transition temperatures ($T_g$) and good morphological stabilities.[34–39] The two TPA groups were positioned across the fluorene bridge to mediate strong electronic coupling between the two redox-active units (Fig. S4). This strong electronic coupling leads to a 0.32-V difference between the first and second TPA⁺·/TPA reduction potentials that raises the HOMO energy to -5.27 eV. This value that maintains a high cell voltage while still accommodating hole extraction from the perovskite with a valence band maximum of -5.4 eV.[40] The high LUMO energy of -2.29 eV maintains a large optical band gap of 2.98 eV to avoid the undesirable absorption of visible light (Fig. S4).

A powerful feature of **DFH** is the propensity for the molecules, upon annealing, to dimerize and preferentially order orthogonal to the substrate, but without additional long-range order in the film. This molecular ordering therefore presents the opportunity to balance the competing need to provide sufficient structural order for charge transport,[41–43] while avoiding the formation of large crystalline domains that introduce cracks or pinholes and impair interfacial electronic contacts.[44,45] This axial growth of **DFH** is largely a consequence of the cyclic, polar 1,3-dioxolane group (Fig. 3a) facilitating strong intermolecular interactions to form tightly bound dimers of **DFH**. A single-crystal structure determination of **DFH** confirms these intermolecular C-H···O and C-H···π interactions between the dioxane groups on neighboring **DFH** molecules parallel to the *b*-axis (Fig. 3b and Fig. S5-6). The strengths of these intermolecular interactions are calculated to be as strong as 110 kJ/mol per pair of **DFH** (ESI). Amorphous **DFH** transforms into paracrystalline dimers when **DFH** is heated above the glass transition temperature ($T_g$) of ~120 °C (Fig. 3c). Slow crystallization occurs at higher annealing temperatures below the onset temperature of cold-crystallization ~temperature ($T_{cc}$) of 160 °C where heterogeneous nucleation becomes operative. Solid-state NMR spectroscopic experiments were employed to track the chemical environment about the carbon atoms of **DFH** upon heating. It is evident that heating amorphous **DFH** to 120 °C causes the ethylene ¹³C resonance of the 1,3-dioxane group to sharpen towards a full-width-half-maximum (FWHM) approaching that of the crystalline material (Figs. 3d, S9 and S10), and the sharpening of the peaks corresponding to crystallization occurs at much higher temperatures

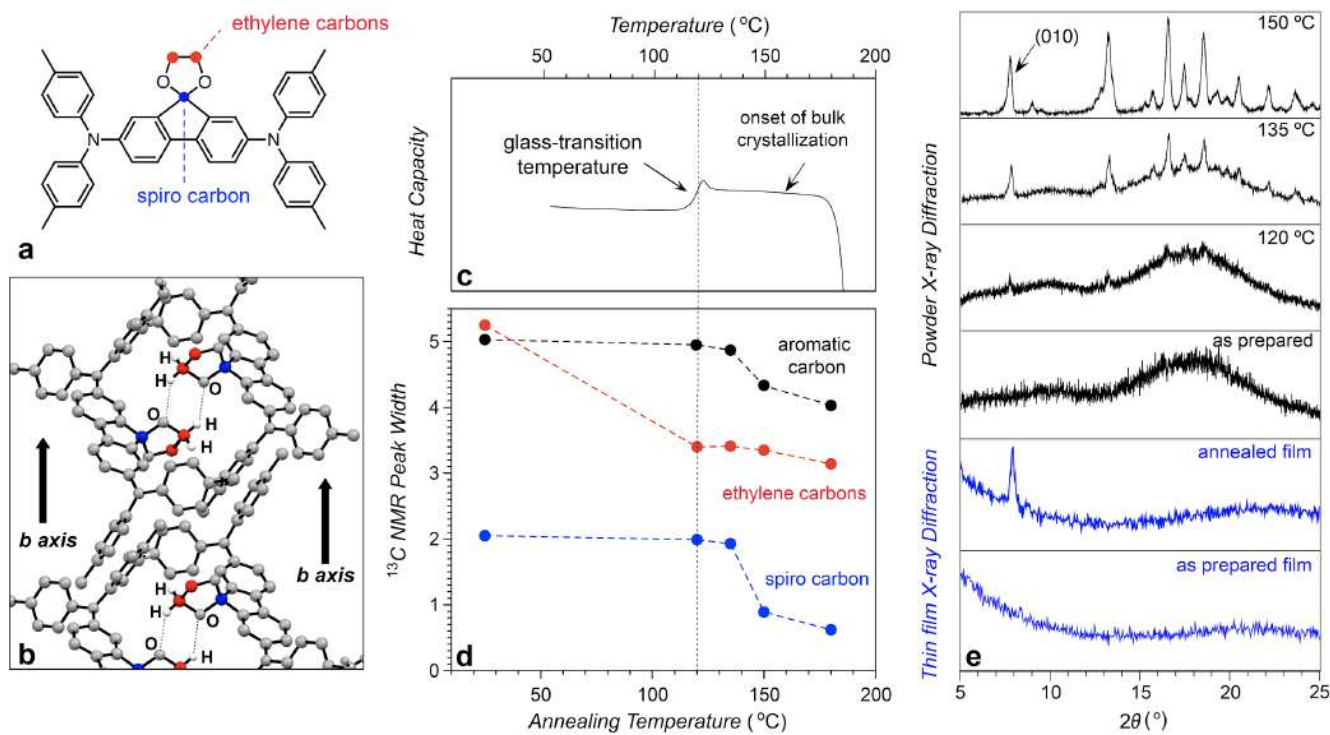

**Fig. 3.** (a) Molecular structure of **DFH** with the ethylene and spiro carbon atoms of the cyclic 1,3-dioxolane group highlighted in red and blue, respectively. (b) The propensity of CH···O interactions between **DFH** molecules illustrated by a ball-and-stick model derived from the X-ray crystallographic data. The spiro carbon and carbon atoms on the ethylene group are highlighted in blue and red, respectively. (c) Differential scanning calorimetry data of amorphous **DFH** powder obtained at a scan rate of 10 K/min. (d) Measured solid-state NMR peak width of selected carbon atoms in terms of *FWHM* of amorphous and annealed **DFH** solids. (e) Powder XRD diffractograms of **DFH** before and after thermal crystallization (black) and GIXD traces of 150 nm spin-coated **DFH** thin films before and after annealing at 150 °C (blue).

Powder XRD data recorded upon heating amorphous **DFH** confirms that crystallization begins to occur at 120 °C, and that higher temperatures yield progressively sharper reflections (Fig. 3e). We then performed *2θ* scans on a 150-nm thin film of **DFH** using parallel-beam grazing-incidence X-ray diffraction (GIXD) and observed the onset only a single peak upon annealing to 150 °C, which is assigned to diffraction signal of the (010) crystallographic planes (Fig. S7). This peak at low *2θ* points to the preferential ordering of **DFH** normal to the plane of the substrate and along the direction of dimerization, with the broad baseline indicating a lack of long-range order for the balance of the film.

An important outcome of the paracrystalline ordering of **DFH** molecules is that the redox-active TPA groups are drawn close to each other without the growth of large crystals. We infer that this feature is responsible for the high hole mobilities ($\mu_h$) of ~1×10$^{-3}$ cm$^2$·V$^{-1}$·s$^{-1}$ for the amorphous and annealed films of **DFH** measured normal to the plane of the substrate. These values are comparable to single-crystal Spiro-OMeTAD,[41] and are greater than those of the previous state-of-the-art dopant-free molecular HTM[20] and heavily p-doped thin films of spiro-OMeTAD.[38,46] The electric conductivity of **DFH** thin films also doubled upon annealing (Table S5). Another notable observation is that we were able to grow pinhole-free perovskite thin-films with large grain sizes (Fig. S14, average 0.6 μm$^2$) on layers of **DFH** annealed at 135 °C, while smaller grains were observed with other **DFH** layers (Fig. S15). Grazing incidence X-ray diffraction (GIXD) of the perovskite layer showed slightly higher tendency of out-plane growth on **DFH** annealed at 135 °C, compared to as-prepared and 120 °C annealed **DFH** (Fig. 16). It is possible that a low degree of **DFH** crystallization at 135 °C is crucial for the growth of large perovskite grains whereas its impact on perovskite crystal orientation is relatively small.

This combination of high axial $\mu_h$ for **DFH** and large grain sizes of the light-absorbing layer provides the ideal conditions for high PSC performance. We therefore tested **DFH** in a planar inverted (p-i-n) PSC with an ITO|**DFH**|perovskite|C$_{60}$|BCP|Ag configuration (where ITO = indium-doped tin oxide; perovskite = MA$_{0.9}$FA$_{0.1}$PbI$_{3-x}$Cl$_x$; BCP = bathocuproine). A mixed methylammonium (MA) / formamidinium perovskite was deposited onto spin-coated **DFH** layers using established methods.[47] Devices containing as-prepared **DFH** generated *PCE* values of merely 10%. This output was increased up to 20.6% when the **DFH** films were annealed at 135°C (Fig. 4a). This *PCE* represents the highest value ever reported for a dopant-free molecular HTM, and the highest value for a device with any dopant-free HTM without interlayers. This *PCE* also compared favorably to the *PCE*s that we were able to produce in optimized devices containing the champion dopant-free polymeric HTM, poly[bis(4-phenyl)(2,4,6-trimethylphenyl)amine] (PTAA; Table 1, Figure S11).[9]

**Table 1.** Tabulated parameters of PSC based on **DFH** treated under different conditions.

| HTM and treatment | $J_{sc}$ (mA·cm$^{-2}$) | $V_{oc}$ (V) | FF | PCE (%) | best PCE (%) |
|---|---|---|---|---|---|
| **DFH** (as prepared) | 18.9 ± 0.8 | 0.86 ± 0.06 | 0.52 ± 0.03 | 8.4 ± 1.1 | 10.2 |
| **DFH** (annealed @ 120 °C) | 22.6 ± 0.1 | 1.09 ± 0.01 | 0.70 ± 0.01 | 17.2 ± 0.3 | 17.6 |
| **DFH** (annealed @ 135 °C) | 22.0 ± 0.7 | 1.08 ± 0.01 | 0.81 ± 0.02 | 19.3 ± 0.7 | 20.6 |
| **DFH** (annealed @ 150 °C) | 22.4 ± 0.3 | 1.04 ± 0.01 | 0.82 ± 0.01 | 19.2 ± 0.5 | 19.7 |
| **PTAA** (annealed @ 100 °C)[a] | 22.4 ± 0.1 | 1.09 ± 0.01 | 0.78 ± 0.02 | 19.0 ± 0.3 | 19.2 |
| **KR321**[b] | 20.9 ± 0.5 | 1.11 ± 0.05 | 0.76 ± 0.03 | 17.7 ± 0.4 | 19.0 |

[a] Our measurements recorded on previously reported HTM at optimized condition.[9]

[b] Data from the previously reported champion device incorporating dopant-free molecular HTM.[20]

The PSC efficiency was doubled by annealing amorphous **DFH** films to temperatures between $T_g$ and $T_{cc}$ to encourage dimerization and a small degree of crystallization. The best devices we tested were those where **DFH** was annealed at 135 °C (Fig. 4a). An annealed **DFH** film quenched >95% of the baseline photoluminescence of the perovskite, indicating a band alignment that enables effective extraction of holes from the perovskite layer (Fig. 4b). The temporal resolution of the perovskite photoluminescence shows that 99% of holes are extracted within 20 ns by annealed **DFH** (Fig. 4c). The same set of steady-state and kinetic experiments recorded with amorphous **DFH** films showed relatively inferior hole extraction properties: ~60% quenching of steady state photoluminescence and >40 ns to extract 99% of holes. We assume that the large defect-free grain sizes play a critical role in mitigating ohmic losses and producing the high measured fill factor (FF) and open-circuit voltage ($V_{oc}$) values.[19,40,48–50]

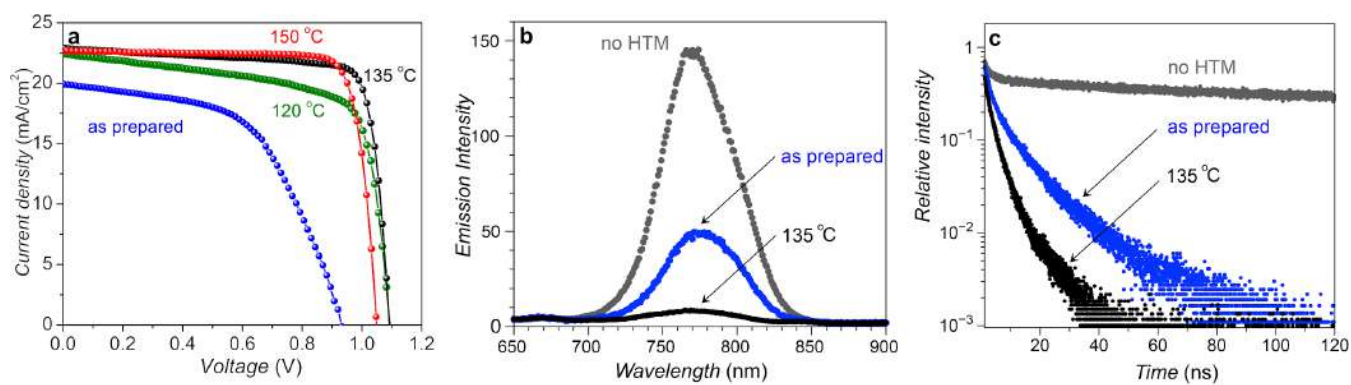

**Fig. 4.** (a) *J-V* curves of PSCs based on **DFH** hole transport layer annealed at different temperatures. (b) steady-state fluorescence quenching of perovskite thin films by the underlying **DFH** layers; (c) fluorescence decay kinetics of PMMA encapsulated perovskite thin films on different **DFH** thin films illuminated by a 553-nm laser.

## Conclusions

We have demonstrated that a molecular organic HTM with redox-active TPA groups yields electronic properties conducive to high performance in an inverted PSC without the need for dopants. The ability to produce a *PCE* >20% with a synthetically accessible HTM and a simple device structure (i.e., no interlayers or barrier layers) offers the opportunity for scalable manufacturing. This record device performance for a dopant-free molecular HTM is a result of the molecular organization of HTM molecules with carefully tailored energy levels, as well as the large grain sizes of the perovskite layer grown on the HTM. This work also shows how thermal annealing of HTMs capable of supramolecular interactions can also affect the quality of the perovskite layers to further drive up device performance. The success with **DFH** also challenges the accepted dogma that dopant-free HTMs need to exhibit large, planar π-stacking structures or complex donor-acceptor moieties.

## Conflicts of Interest

There are no conflicts to declare

## Notes and References:

1    M. M. Lee, J. Teuscher, T. Miyasaka, T. N. Murakami and H. J. Snaith, *Science*, 2012, **338**, 643–647.

2    J. H. Heo, S. H. Im, J. H. Noh, T. N. Mandal, C.-S. Lim, J. A. Chang, Y. H. Lee, H.-J. Kim, A. Sarkar, M. K. Nazeeruddin, M. Grätzel and S. I. Seok, *Nat. Photonics*, 2013, **7**, 486.

3    J. Burschka, N. Pellet, S.-J. Moon, R. Humphry-Baker, P. Gao, M. K. Nazeeruddin and M. Grätzel, *Nature*,

2013, **499**, 316–319.

4   H. Zhou, Q. Chen, G. Li, S. Luo, T.-B. Song, H.-S. Duan, Z. Hong, J. You, Y. Liu and Y. Yang, *Science*, 2014, **345**, 542–546.

5   P. Kung, M. Li, P. Lin, Y. Chiang, C. Chan, T. Guo and P. Chen, *Adv. Mater. Interfaces*, 2018, **5**, 1800882.

6   J. Lim, S. Y. Kong and Y. J. Yun, *J. Nanomater.*, 2018, 7545914.

7   C. H. Teh, R. Daik, E. L. Lim, C.-C. Yap, M. A. Ibrahim, N. A. Ludin, K. Sopian and M. A. M. Teridi, *J. Mater. Chem. A.*, 2016, **4**, 15788–15822.

8   T. P. Osedach, T. L. Andrew and V. Bulović, *Energy Environ. Sci.*, 2013, **6**, 711–718.

9   M. Stolterfoht, C. M. Wolff, Y. Amir, A. Paulke, L. Perdigón-Toro, P. Caprioglio and D. Neher, *Energy Environ. Sci.*, 2017, **10**, 1530–1539.

10  Q. Jiang, Y. Zhao, X. Zhang, X. Yang, Y. Chen, Z. Chu, Q. Ye, X. Li, Z. Yin and J. You, *Nat. Photonics*, 2019, **13**, 500.

11  N. J. Jeon, H. Na, E. H. Jung, T.-Y. Yang, Y. G. Lee, G. Kim, H.-W. Shin, S. Il Seok, J. Lee and J. Seo, *Nature Energy*, 2018, **3**, 682–689.

12  T. H. Schloemer, J. A. Christians, J. M. Luther and A. Sellinger, *Chem. Sci.*, 2019, **10**, 1904–1935.

13  J. A. Christians. P. A. Miranda Herrera and P. V. Kamat, *J. Am. Chem. Soc.*, 2015, **137**, 1530–1538.

14  J. Yang and T. L. Kelly, *Inorg. Chem.*, 2017, **56**, 92–101.

15  Z. Li, C. Xiao, Y. Yang, S. P. Harvey, D. H. Kim, J. A. Christians, M. Yang, P. Schulz, S. U. Nanayakkara, C.-S. Jiang, J. M. Luther, J. J. Berry, M. C. Beard, M. M. Al-Jassim and K. Zhu, *Energy Environ. Sci.*, 2017, **10**, 1234–1242.

16  J. Zhang, Q. Daniel, T. Zhang, X. Wen, B. Xu, L. Sun, U. Bach and Y.-B. Cheng, *ACS Nano*, 2018, **12**, 10452–10462.

17  F. Wang, A. Shimazaki, F. Yang, K. Kanahashi, K. Matsuki, Y. Miyauchi, T. Takenobu, A. Wakamiya, Y. Murata and K. Matsuda, *J. Phys. Chem. C*, 2017, **121**, 1562–1568.

18  A. Agresti, S. Pescetelli, B. Taheri, A. E. Del Rio Castillo, F. Cinà, F. Bonaccorso and A. Di Carlo, *ChemSusChem*, 2016, **9**, 2609–2619.

19  C. Huang, W. Fu, C.-Z. Li, Z. Zhang, W. Qiu, M. Shi, P. Heremans, A. K.-Y. Jen and H. Chen, *J. Am. Chem. Soc.*, 2016, **138**, 2528–2531.

20  K. Rakstys, S. Paek, P. Gao, P. Gratia, T. Marszalek, G. Grancini, K. T. Cho, K. Genevicius, V. Jankauskas, W. Pisula and M. K. Nazeeruddin, *J. Mater. Chem. A*, 2017, **5**, 7811–7815.

21  S. Paek, P. Qin, Y. Lee, K. T. Cho, P. Gao, G. Grancini, E. Oveisi, P. Gratia, K. Rakstys, S. A. Al-Muhtaseb, C. Ludwig, J. Ko and M. K. Nazeeruddin, *Adv. Mater.*, 2017, **29**, 4706–4713.

22  R. Azmi, S. Y. Nam, S. Sinaga, Z. A. Akbar, C.-L. Lee, S. C. Yoon, I. H. Jung and S.-Y. Jang, *Nano Energy*, 2018, **44**, 191–198.

23  M. Cheng, B. Xu, C. Chen, X. Yang, F. Zhang, Q. Tan, Y. Hua, L. Kloo and L. Sun, *Adv. Energy Mater.*, 2015, **5**, 1401720.

24  Y. Liu, Q. Chen, H.-S. Duan, H. Zhou, Y. (michael) Yang, H. Chen, S. Luo, T.-B. Song, L. Dou, Z. Hong and Y. Yang, *J. Mater. Chem. A.*, 2015, **3**, 11940–11947.

25  Z. Li, Z. Zhu, C.-C. Chueh, S. B. Jo, J. Luo, S.-H. Jang and A. K.-Y. Jen, *J. Am. Chem. Soc.*, 2016, **138**, 11833–11839.

26  D. Bi, A. Mishra, P. Gao, M. Franckevičius, C. Steck, S. M. Zakeeruddin, M. K. Nazeeruddin, P. Bäuerle, M. Grätzel and A. Hagfeldt, *ChemSusChem*, 2016, **9**, 433–438.

27  E. H. Jung, N. J. Jeon, E. Y. Park, C. S. Moon, T. J. Shin, T.-Y. Yang, J. H. Noh and J. Seo, *Nature*, 2019, **567**, 511–515.

28  E. Rezaee, X. Liu, Q. Hu, L. Dong, Q. Chen, J.-H. Pan and Z.-X. Xu, *Sol. RRL*, 2018, **2**, 1800200.

29  W. Zhou, Z. Wen and P. Gao, *Adv. Energy Mater.*, 2018, **8**, 1702512.

30  W. Chen, Y. Wu, Y. Yue, J. Liu, W. Zhang, X. Yang, H. Chen, E. Bi, I. Ashraful, M. Grätzel and L. Han, *Science*, 2015, **350**, 944–948.

31  F. Xie, C.-C. Chen, Y. Wu, X. Li, M. Cai, X. Liu, X. Yang and L. Han, *Energy Environ. Sci.*, 2017, **10**, 1942–1949.

32  S. Mülhopt, S. Diabaté, M. Dilger, C. Adelhelm, C. Anderlohr, T. Bergfeldt, J. Gómez de la Torre, Y. Jiang, E. Valsami-Jones, D. Langevin, I. Lynch, E. Mahon, I. Nelissen, J. Piella, V. Puntes, S. Ray, R. Schneider, T. Wilkins, C. Weiss and H.-R. Paur, *Nanomaterials*, 2018, **8**, 311.

33   H. K. H. Lee, Z. Li, I. Constantinou, F. So, S. W. Tsang and S. K. So, *Adv. Energy Mater.*, 2014, **4**, 1400768.

34   P. Ganesan, K. Fu, P. Gao, I. Raabe, K. Schenk, R. Scopelliti, J. Luo, L. H. Wong, M. Grätzel and M. K. Nazeeruddin, *Energy Environ. Sci.*, 2015, **8**, 1986–1991.

35   M. Saliba, S. Orlandi, T. Matsui, S. Aghazada, M. Cavazzini, J.-P. Correa-Baena, P. Gao, R. Scopelliti, E. Mosconi, K.-H. Dahmen, F. De Angelis, A. Abate, A. Hagfeldt, G. Pozzi, M. Graetzel and M. K. Nazeeruddin, *Nature Energy*, 2016, **1**, 15017.

36   B. Xu, D. Bi, Y. Hua, P. Liu, M. Cheng, M. Grätzel, L. Kloo, A. Hagfeldt and L. Sun, *Energy Environ. Sci.*, 2016, **9**, 873–877.

37   B. Xu, J. Zhang, Y. Hua, P. Liu, L. Wang, C. Ruan, Y. Li, G. Boschloo, E. M. J. Johansson, L. Kloo, A. Hagfeldt, A. K.-Y. Jen and L. Sun, *Chem*, 2017, **2**, 676–687.

38   V. A. Chiykowski, Y. Cao, H. Tan, D. P. Tabor, E. H. Sargent, A. Aspuru-Guzik and C. P. Berlinguette, *Angew. Chem. Int. Ed Engl.*, 2018, **57**, 15529–15533.

39   X. Wang, J. Zhang, S. Yu, W. Yu, P. Fu, X. Liu, D. Tu, X. Guo and C. Li, *Angew. Chem. Int. Ed Engl.*, 2018, **57**, 12529–12533.

40   L. Yang, F. Cai, Y. Yan, J. Li, D. Liu, A. J. Pearson and T. Wang, *Adv. Funct. Mater.*, 2017, **27**, 1702613.

41   D. Shi, X. Qin, Y. Li, Y. He, C. Zhong, J. Pan, H. Dong, W. Xu, T. Li, W. Hu, J.-L. Brédas and O. M. Bakr, *Sci Adv*, 2016, **2**, e1501491.

42   I. Yavuz and K. N. Houk, *J. Phys. Chem. C*, 2017, **121**, 993–999.

43   R. Noriega, J. Rivnay, K. Vandewal, F. P. V. Koch, N. Stingelin, P. Smith, M. F. Toney and A. Salleo, *Nat. Mater.*, 2013, **12**, 1038–1044.

44   Y. Fang, X. Wang, Q. Wang, J. Huang and T. Wu, *Phys. Status Solidi* , 2014, **211**, 2809–2816.

45   T. Malinauskas, D. Tomkute-Luksiene, R. Sens, M. Daskeviciene, R. Send, H. Wonneberger, V. Jankauskas, I. Bruder and V. Getautis, *ACS Appl. Mater. Interfaces*, 2015, **7**, 11107–11116.

46   J. Luo, J. Xia, H. Yang, L. Chen, Z. Wan, F. Han, H. A. Malik, X. Zhu and C. Jia, *Energy Environ. Sci.*, 2018, **11**, 2035–2045.

47   Y. Li, W. Sun, F. Gu, D. Ouyang, Z. Bian, Z. Liu, W. C. H. Choy and T. L. Kelly, *Adv. Mater. Interfaces*, in press.

48   J. C. Yu, J. A. Hong, E. D. Jung, D. B. Kim, S.-M. Baek, S. Lee, S. Cho, S. S. Park, K. J. Choi and M. H. Song, *Sci. Rep.*, 2018, **8**, 1070.

49   W. Chen, Y. Wu, J. Fan, A. B. Djurišić, F. Liu, H. W. Tam, A. Ng, C. Surya, W. K. Chan, D. Wang and Z.-B. He, *Adv. Energy Mater.*, 2018, **8**, 1703519.

50   D. Bi, C. Yi, J. Luo, J.-D. Décoppet, F. Zhang, S. M. Zakeeruddin, X. Li, A. Hagfeldt and M. Grätzel, *Nature Energy*, 2016, **1**, 16142.

**Acknowledgments**

The authors are grateful to the Canadian Natural Science and Engineering Research Council (RGPIN 337345-13; RGPIN-2017-03732), Canadian Foundation for Innovation (229288), Canadian Institute for Advanced Research (BSE-BERL-162173), the University of Saskatchewan and Canada Research Chairs for financial support. This research was undertaken thanks in part to funding from the Canada First Research Excellence Fund, Quantum Materials and Future Technologies Program. SEM imaging was performed in the Centre for High-Throughput Phenogenomics at the University of British Columbia, a facility supported by the Canada Foundation for Innovation, British Columbia Knowledge Development Foundation, and the UBC Faculty of Dentistry.

Supplementary Materials for

# Dopant-free molecular hole transport material that mediates a 20% power conversion efficiency in a perovskite solar cell

Yang Cao[1,4,†], Yunlong Li[2,†], Thomas Morrissey[1,4],  Brian Lam[1], Brian O. Patrick[1], David J. Dvorak[4], Zhicheng Xia[1], Timothy L. Kelly[2,*], Curtis P. Berlinguette[1,3,4,*]

correspondence to tim.kelly@usask.ca and cberling@chem.ubc.ca
†These authors contributed equally to this work

## Synthetic Methods

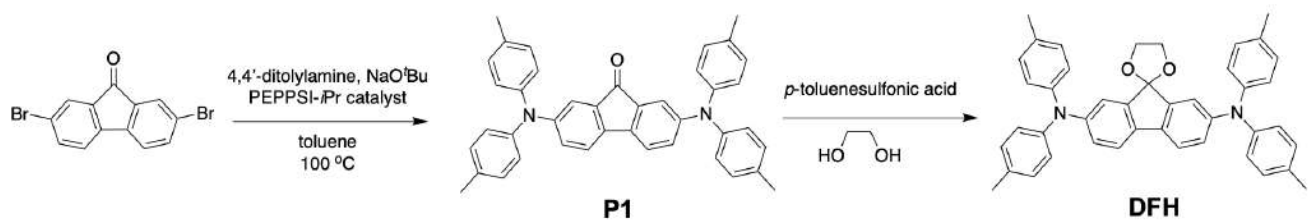

**Fig. S1.** Synthesis of compounds **P1** and **DFH**.

9-Fluorenone (98.0%, TCI America), 2,7-dibromo-9-fluorenone (98.0%, TCI America), *p*,*p'*-ditolylamine (98.0%, TCI America), ethylene glycol (anhydrous, 99.8%, Sigma-Aldrich), 1,2-propanediol (99.5%, Sigma-Aldrich), *p*-toluenesulfonic acid monohydrate (99%, Sigma-Aldrich), sodium *tert*-butoxide (97%, Sigma-Aldrich) and PEPPSI-IPr catalyst (98%, Sigma-Aldrich) were purchased and directly used without purification. All synthesis steps were performed using anhydrous toluene from a solvent purification system. $^1$H and $^{13}$C{$^1$H} NMR spectra were collected on a Bruker AV400DIR instrument at 25.0 °C with resonance frequencies of 400 MHz and 101 MHz for $^1$H and $^{13}$C nuclei, respectively. Chemical shifts (δ) are reported in parts per million (ppm) using the protio-solvent signals δ 7.26 and 77.0 for chloroform-*d*, δ 5.32 and 53.84 for methylene chloride-$d_2$ for $^1$H and $^{13}$C NMR spectra, respectively. Standard abbreviations indicating multiplicity are used as follows: s = singlet; d = doublet; dd = doublet of doublets; m = multiplet.

Synthesis of **P1**: The synthesis of **P1** was carried out under $N_2$ using standard Schlenk techniques. An oven-dried 100 mL two-neck round-bottom flask was connected to the Schlenk line on one neck by via a condenser. A rubber septum was attached to the other neck and the flask was cycled three times between vacuum and nitrogen. 2,7-Dibromo-9-fluorenone (3.38 g, 10.0 mmol), *p*,*p'*-ditolylamine (4.14 g, 21.0 mmol), sodium *tert*-butoxide (2.02 g, 21.0 mmol), PEPPSI™-IPr catalyst (68 mg, 0.10 mmol) and a stir bar were added to the flask and quickly cycled three times. Anhydrous toluene (40 mL) was added via a syringe through the rubber septum which was replaced with a glass stopper under high nitrogen flow. The mixture was stirred for 5 min at room temperature before it was warmed up to 100 °C using an oil bath. After 18 hours, the mixture was poured into 250 mL of methanol and sonicated for 20 min. The dark precipitate was collected via vacuum filtration and washed with 20 mL 0.1M HCl solution, 20 mL deionized water and twice with 20 mL methanol. The purple powder was recrystallized from ethanol/toluene and 5.10 g of pure product was obtained (89 % yield). $^1$H NMR (400 MHz, chloroform-*d*) δ 7.29 (d, *J* = 2.2 Hz, 2H), 7.19 (d, *J* = 8.2 Hz, 2H), 7.14 – 7.04 (m, 10H), 7.00 (d, *J* = 8.4 Hz, 8H), 2.34 (s, 12H). $^{13}$C{$^1$H} NMR (101 MHz, chloroform-*d*) δ 193.90, 148.63, 144.86, 137.53, 135.81, 133.20, 130.20, 127.48, 124.91, 120.23, 118.50, 20.96. HRMS (ESI): *m/z* = 570.2675 [M$^+$] (calcd for [C$_{41}$H$_{34}$ON$_2$]$^+$: *m/z* = 570.2671).

Synthesis of **DFH**: **P1** (2.00 g, 3.50 mmol), *p*-toluenesulfonic acid monohydrate (50 mg, 0.26 mmol), ethylene glycol (2.00 mL, 32.2 mmol) and a stir bar were added to a round bottom flask which was connected with a condenser via a Dean-Stark trap. The flask was cycled three times between vacuum and $N_2$ before 40 mL anhydrous toluene was added via syringe. The reaction mixture was heated at 100 °C under $N_2$ for 2 days before it was concentrated to around 20 mL under vacuum and poured into 200 mL of methanol. The light yellow precipitate (1.85 g, 84% yield) was collected after vacuum filtration, washing with methanol and drying under high vacuum. After filtration, rinsing and drying under high vacuum 1.77 g (82 % yield) of **DFH** was obtained as a pale white powder. $^1$H NMR (400 MHz, methylene chloride-$d_2$) δ 7.29 (d, $J$ = 8.2 Hz, 2H), 7.12 – 7.04 (m, 10H), 6.98 (d, $J$ = 8.4 Hz, 8H), 6.94 (dd, $J$ = 8.2, 2.1 Hz, 2H), 4.12 (s, 4H), 2.31 (s, 12H). $^{13}$C{$^1$H} NMR (101 MHz, methylene chloride-$d_2$) δ 148.36, 146.08, 145.87, 133.81, 133.13, 130.41, 125.35, 124.90, 120.37, 119.11, 112.39, 66.16, 21.07. HRMS (ESI): $m/z$ = 614.2937 [M$^+$] (calcd for [C$_{43}$H$_{38}$O$_2$N$_2$]$^+$: $m/z$ = 614.2933).

Synthesis of **P2**: P2 without triphenyl amine groups were made to help with peak assignments in ssNMR. 9-Fluorenone (1.26 g, 7.00mmol), *p*-toluenesulfonic acid monohydrate (100 mg, 0.52 mmol), ethylene glycol (4.00 mL, 64.4 mmol) and a stir bar were added to a round bottom flask which was connected with a condenser via a Dean-Stark trap. The reaction mixture was heated at 100 °C under N$_2$ for 2 days before it was concentrated to around 10 mL under vacuum and poured into 100 mL of methanol. The off-white precipitate (1.22 g, 78% yield) was collected after vacuum filtration, washed with methanol and dried under high vacuum. $^1$H NMR (400 MHz, chloroform-*d*) δ 7.58 (dt, $J$ = 7.5, 0.9 Hz, 2H), 7.47 (dt, $J$ = 7.4, 1.0 Hz, 2H), 7.39 (td, $J$ = 7.5, 1.2 Hz, 2H), 7.29 (td, $J$ = 7.5, 1.1 Hz, 2H), 4.43 (s, 4H). $^{13}$C{$^1$H} NMR (101 MHz, chloroform-*d*) δ 144.24, 139.75, 130.30, 128.36, 123.84, 120.04, 112.51, 65.94. HRMS (ESI): $m/z$ = 225.0919 [M+H$^+$] (calcd for [C$_{43}$H$_{38}$O$_2$N$_2$]$^+$: $m/z$ = 225.0916).

## Cost Analysis

The material cost was calculated using the lowest available price from the same chemical vendor used in the laboratory gram-scale synthesis (**Table S1** and **S2**).

**Table S1.** Material Cost to Synthesize **P1**

| Chemical | Unit price/cost | Amount used | Cost |
|---|---|---|---|
| 2,7-Dibromo-9-fluorenone | 73 USD per 25 g (TCI) | 3.38 g | 9.87 USD |
| Sodium *tert*-butoxide | 491 USD per 1.5 kg (Sigma) | 2.02 g | 0.66 USD |
| *p,p'*-Ditolylamine | 70 USD per 25 g (TCI) | 4.14 g | 11.59 USD |
| PEPPSI™-IPr catalyst | 10,548.83 USD per 250 g (Sigma) | 68 mg | 2.87 USD |
| Toluene | 517.67 USD per 20 L (Sigma) | 40 mL | 1.04 USD |
| **P1** | **5.10 USD per 1 g** | **5.10 g** | **26.03 USD** |

**Table S2.** Material cost to synthesize **DFH**

| Chemical | Unit price/cost | Amount used | Cost |
|---|---|---|---|
| **P1** | 5.10 USD per 1 g | 2.00 g | 10.20 USD |
| Ethylene glycol | 341.74 USD per 6 L (Sigma) | 2.00 mL | 0.11 USD |
| Toluene | 517.67 USD per 20 L (Sigma) | 40 mL | 1.04 USD |
| *p*-Toluenesulfonic acid monohydrate | 101.16 per 2.5 kg | 50 mg | 0.002 USD |
| **DFH** | **6.41 USD per 1g** | **1.77 g** | **11.35 USD** |

*Characterization of Compounds*

Solution UV-Vis absorption spectra were collected on a Cary 5000 spectrophotometer (**Fig. S2A**). Solution photoluminescence spectra were recorded with a Cary Eclipse fluorimeter (**Fig. S2B**). All solution samples were measured in a 1 cm quartz cell at room temperature in HPLC grade DCM. The concentrations of the DCM solutions of analytes for UV-Vis and photoluminescence measurements were $2\times10^{-5}$ mol$\cdot$L$^{-1}$ and $1\times10^{-5}$ mol$\cdot$L$^{-1}$, respectively. Solid-state absorptance (%A) was measured on a Cary 7000 spectrophotometer by subtracting reflection (%R) and transmission (%T) from incident light (**Fig. S3**).

Differential pulse voltammetry (DPV) data was recorded with a CHI660D potentiostat at room temperature using a platinum wire counter electrode and a platinum working electrode (**Fig. S4**). Ag/AgCl in saturated KCl was used as the reference electrode and was calibrated versus the normal hydrogen electrode (NHE) by the addition of ferrocene. A 0.1 M *n*-NBu$_4$PF$_6$ electrolyte solution in DCM was used for all HTMs. DPV data were acquired for 0.5 mM solutions of compounds at a scan rate of 50 mV$\cdot$s$^{-1}$. Differential scanning calorimetry (DSC) data was recorded using a Netzsch DSC Polyma 214 calorimeter under a N$_2$ purge flow at a scan rate of 10 K$\cdot$min$^{-1}$.

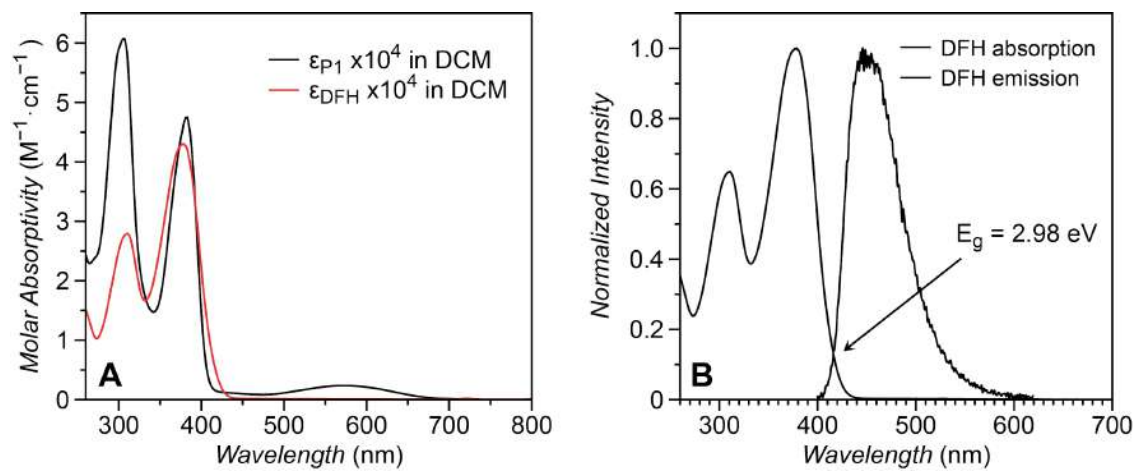

**Fig. S2.** (a) Molar absorptivity of **P1** and **DFH** in dichloromethane. (b) Normalized absorption and emission spectra of **DFH**.

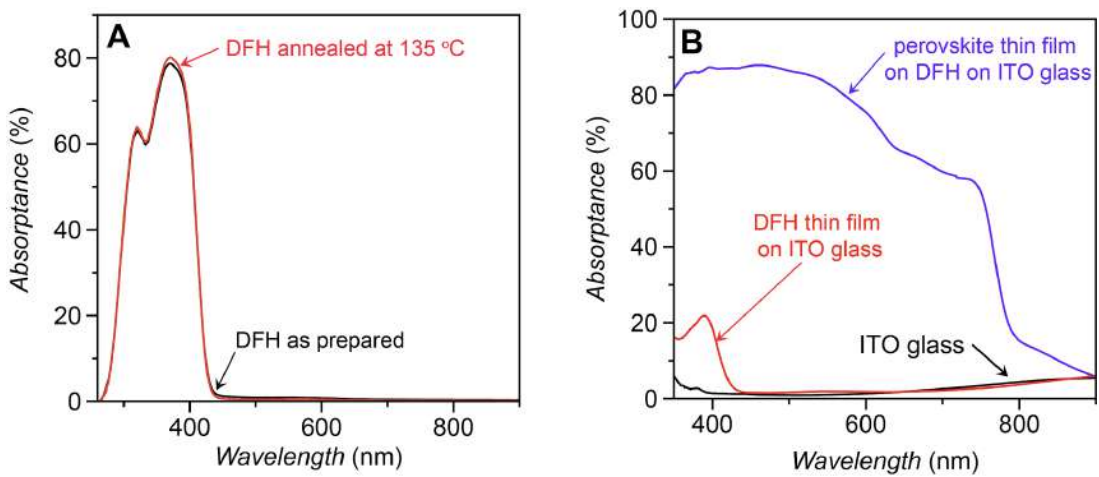

**Fig. S3.** (A) UV-Vis spectra of spin-coated (60 mg/mL in chlorobenzene, 3000 rpm for 30s) solid state thin films of **DFH**, unheated and annealed at 135 °C for 20 mins. (B) UV-Vis spectra of the ITO glass substrate, DFH on ITO glass substrate (spin-coated at 6000 rpm for 30s from a 15 mg/mL solution in chlorobenzene, and annealed at 135 °C for 20 mins), and perovskite thin film (coated with the same method used for the fabrication of full devices) on aforementioned DFH on ITO glass substrate.

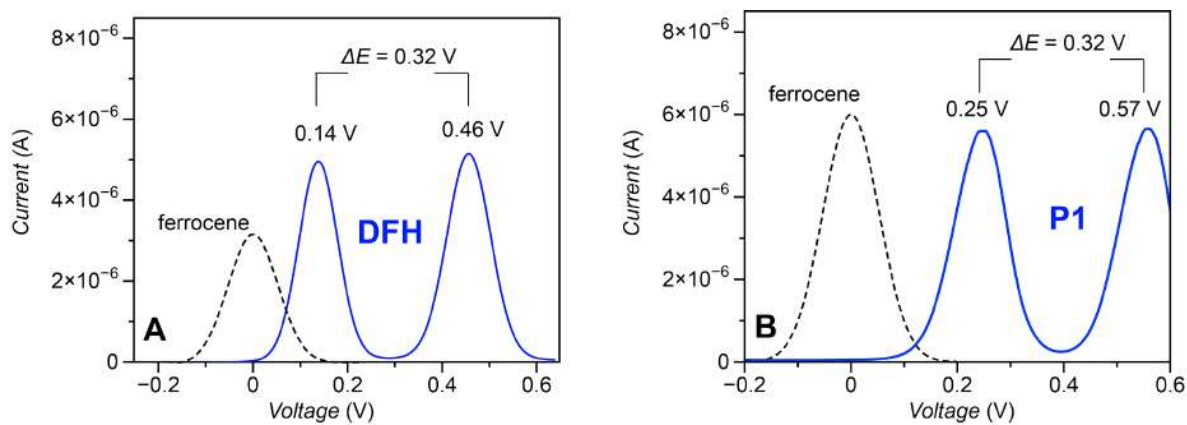

**Fig. S4.** Differential pulse voltammetry (DPV) of (A) **DFH** and (B) **P1** in dichloromethane showing the voltage of two strongly coupled oxidation events relative to the $Fc^+/FcH$ redox couple.

**Table S3.** Summary of the Optical, Electrochemical and Electrical Properties of **DFH**.

| $\lambda_{abs}$ (nm) | $\lambda_{em}$ (nm) | $E_{1/2}$ (V) | $E_{gap}$ (eV) | $E_{HOMO}$ (eV) |
|---|---|---|---|---|
| 370 | 450 | 0.14 | 2.98 | -5.27 |

Half-wave redox potentials ($E_{1/2}$) in DCM are relative to that of ferrocene ($E_{FcH/Fc^+}$ = 630 mV vs NHE). $E_{gap}$ was estimated from the intersection of normalized absorption and emission spectra. $E_{HOMO}$ (eV) = - 4.5 - $E_{1/2}$ (V vs NHE).

## X-ray Crystallography

Single crystal X-ray crystallography was performed using a Bruker APEX II area detector diffractometer. Single colourless tablet-shaped crystals of **DFH** were recrystallized from a mixture of toluene and ethanol by slow diffusion. Suitable crystals were selected and mounted on a mylar loop. Data were measured using MoK$_\alpha$ radiation (microfocus sealed X-ray tube, 50 kV, 0.99 mA). The structure was solved with the **SHELXT** [1] structure solution program using the Intrinsic Phasing solution method and by using **Olex2** [2] as the graphical interface. The diffraction pattern indexing, unit cell refinement, data reduction, scaling and absorption corrections were performed using **SAINT** (Bruker, V8.38A, after 2013). Multi-Scan absorption correction was performed using **SADABS**-2016/2 (Bruker, 2016/2). All non-hydrogen atoms were refined anisotropically. Hydrogen atom positions were calculated geometrically and refined using the riding model. **DFH**: The maximum resolution that was achieved was $Q = 30.629°$ (0.70 Å). The structure was solved and the space group $P$-1 (# 2) determined by the **XT** structure solution program using Intrinsic Phasing and refined by Least Squares using version 2017/1 of **XL**. There is a small amount of disorder in the 1,3-dioxolane moiety that was modelled in two orientations.

**Table S4.** Summary of the Structure Parameters of **DFH** Single Crystal.

| | |
|---|---|
| Chemical Formula | $C_{43}H_{38}N_2O_2$ |
| Density of Crystal, $D_{calc.}$ (g·cm$^{-3}$) | 1.220 |
| Formula Weight (g/mol) | 614.75 |
| Colour, Shape and Size of Crystal (mm$^3$) | colourless, tablet, 0.41×0.24×0.12 |
| Crystal System, Space Group | triclinic, $P$-1 |
| $a, b, c$ (Å) | 10.5358(10), 12.5714(12), 15.0206(14) |
| $\alpha, \beta, \gamma$ (°) | 101.715(2), 107.958(2), 109.766(2) |
| V (Å$^3$) | 1672.8(3) |
| Numbers of molecules in unit cell, $Z$ | 2 |

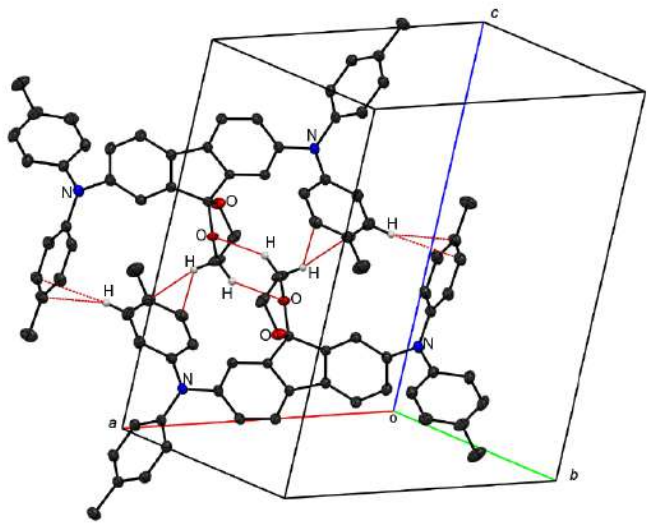

**Fig. S5.** ORTEP thermal ellipsoid presentation of the CH···O hydrogen bonding and C-H···π interactions between dimeric **DFH** molecules. Non-interacting hydrogens have been omitted for clarity.

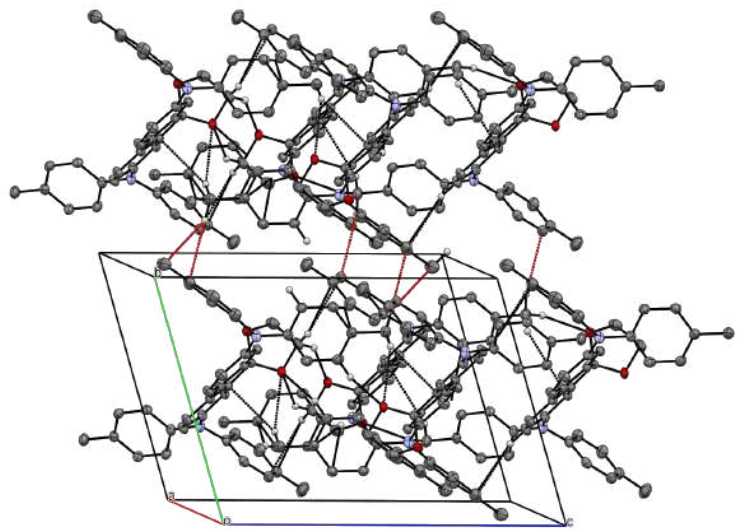

**Fig. S6.** ORTEP thermal ellipsoid presentation of the C-H···π interactions between adjacent (010) face of the **DFH** crystal lattice. Non-interacting hydrogens have been omitted for clarity.

### *Density Functional Theory Calculations*

DFT calculations were carried out using the Gaussian 09 Rev-D.01 software [3]. Molecular properties such as molecular energies, HOMO levels and electrostatic surface potential mapping were modeled using long-range corrected ω-B97XD functional [4] with 6-311G** basis set. The reorganization energy $\lambda$ (0.021 Hartree or 0.57 eV) of **DFH** was calculated the following equation:

$$\lambda = E^{+*} - E - E^{+} + E^{*},$$

where $E$ is the energy of the neutral state with optimized neutral geometry (-1921.2364792 Hartree), $E^{+*}$ is the energy of the cationic state with the optimized neutral-state geometry (-1921.0093504 Hartree), $E^{+}$ is the energy of the cationic state with optimized cationic geometry (-1921.0194704 Hartree) and $E^{*}$ is the energy of the neutral state with the optimized cationic geometry (-1921.225525 Hartree). All four energies are calculated in vacuum. According to literature [5], the reorganization energy of Spiro-OMeTAD was calculated to be 0.31 eV using the same long-range corrected functional.

The electronic interaction energy ($E_{\text{HTM-HTM}}$) between **DFH** dimers with crystallographic atomic coordinates was calculated using counterpoises and corrected for basis set superposition error (BSSE)[6]. The following two different basis sets were used.

$E_{\text{HTM-HTM}}$ = -25.16 kcal/mol or -105 kJ/mol (6-31+G** for interacting H, 6-31G** for everything else)

$E_{\text{HTM-HTM}}$ = -26.19 kcal/mol or -110 kJ/mol (6-311+G** for interacting H, 6-311G** for everything else)

## X-ray Diffraction

Powder and thin film X-ray diffraction experiments were performed on a Rigaku SmartLab X-ray diffractometer using $CuK_\alpha$ radiation. A Bragg-Brentano $\theta$-$\theta$ geometry was used to probe powder samples annealed at different temperatures. The scanning range was from 5° to 25°, with 0.02° per step. The thin film was characterized using 2θ scans using parallel-beam grazing-incidence X-ray diffraction (GIXD). The HTM film was spin-coated on a glass substrate, and annealed at 150 °C on a hotplate until formation of visible spherulites on the thin film. The scanning range was from 5° to 60°, with 0.04° per step, the incidence angle was 0.7°.

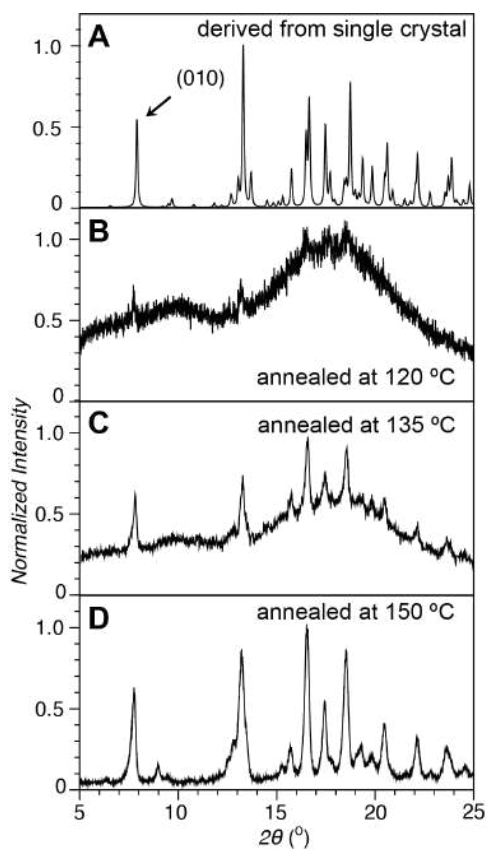

**Fig. S7**. The powder XRD spectra of **DFH** derived from its crystal structure (A) and measured XRD spectra of **DFH** annealed at 120 °C (B), 135 °C (C) and 150 °C (D).

### Solid-State NMR

$^{13}C$ CP-MAS NMR spectra with high power proton decoupling were collected on a 400MHz Bruker solid state DRX spectrometer. Sample was spinning at 6 kHz at magical angle. Ramped pulse on $^{13}C$ frequency was used for cross polarization with a contact time of 4ms for all experiments. Relaxation delay was set to be 5 seconds, and acquisition time 50ms. Data were processed with a 20 Hz line broadening exponential decay function. Chemical shifts (δ, in ppm) were referenced with adamantane $^{13}CH_2$ signal at 29.5ppm. All experiments are performed at room temperature. Peak assignments were carried out using an analogue without the ditolyamine groups (**fig. S7**).

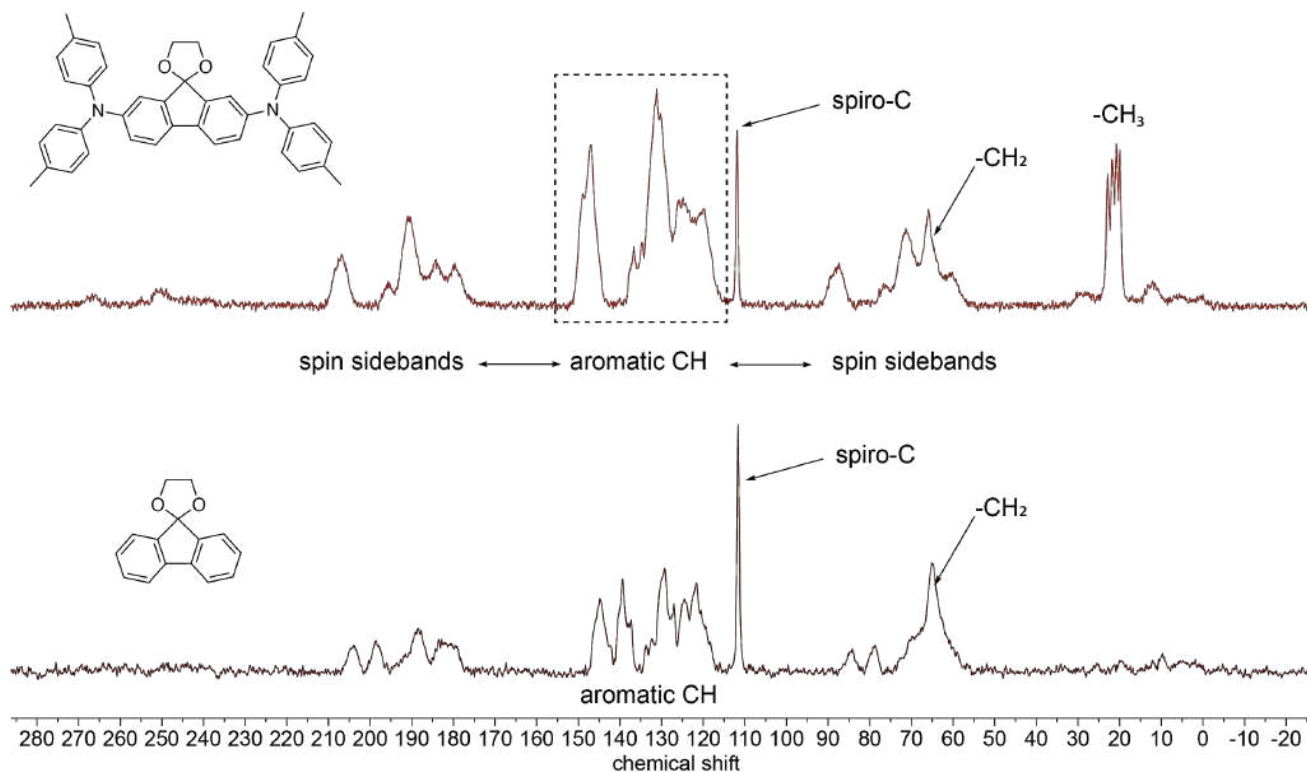

**Fig. S8**. The assignment of peaks and spin-sidebands of crystalline solid-state **DFH** and control **P2**.

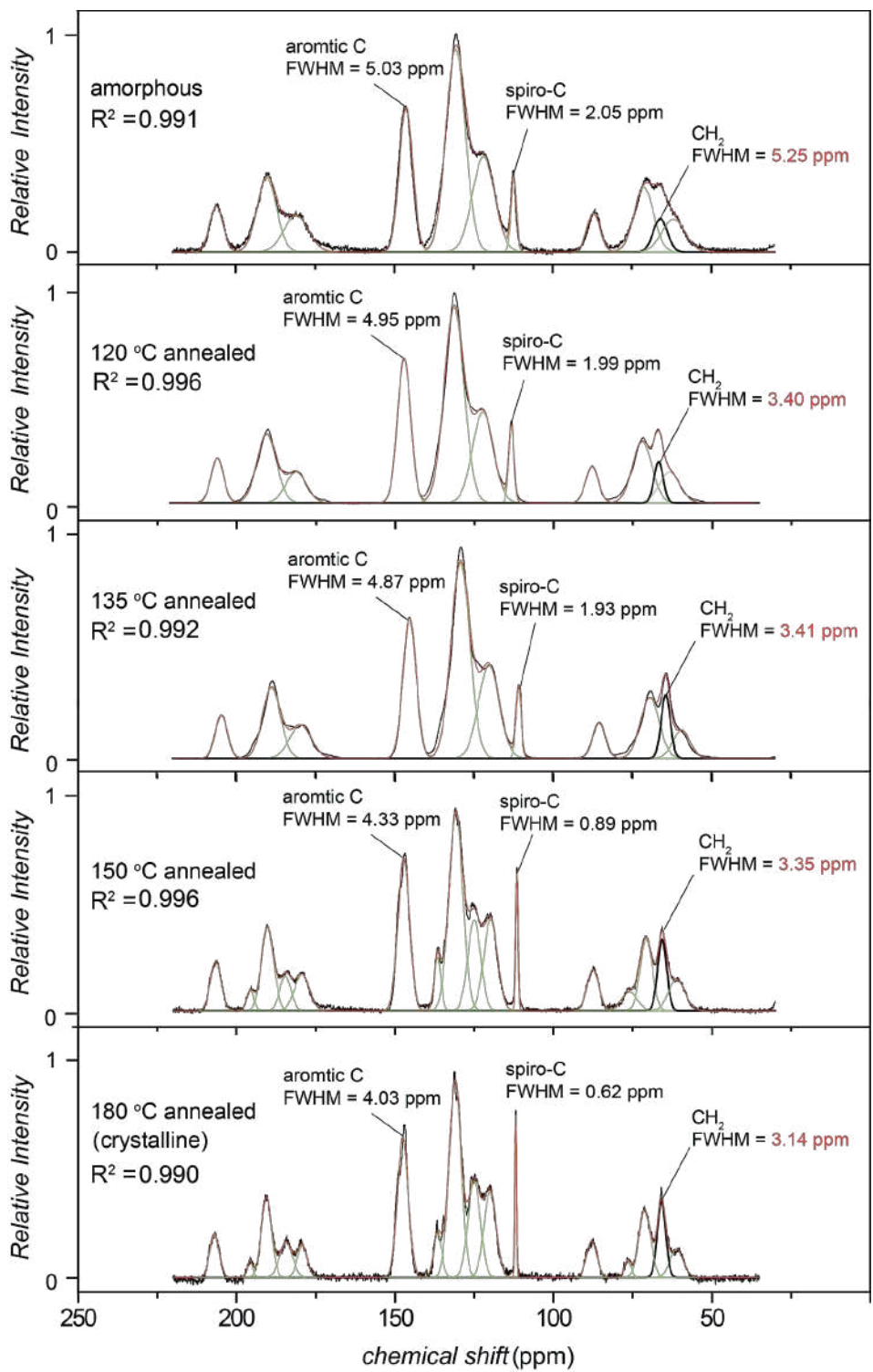

**Fig. S9.** Fitted high power CP-MAS NMR spectra of **DFH** between 220 ppm and 30 ppm, the methylene peak was highlighted in black.

### *Mobility and Conductivity Measurements*

The hole mobility of **DFH** was measured on hole-only devices (ITO/MoO$_3$/HTM/Au) using the space-charge-limited current (SCLC) method. The hole mobility was determined by fitting the quadratic region of the *I/V* curve to the Mott-Gurney law.

$$J_d = \frac{9\varepsilon\varepsilon_0\mu V_b^2}{8L^3},$$

where *L* is the thickness of the **DFH** layer (here $L \approx 40$ nm), the relative dielectric constant of **DFH** (ε) was assumed to be 3, $\varepsilon_0$ is the vacuum permittivity, and $V_b$ is the applied voltage.

Conductivity values were obtained using measurements on four different thin film devices on the same conductive ITO substrate with a spin-coated HTM layer and four sputtered Au contacts (ITO/HTM/Au). The HTM bulk resistance *R* was calculated by subtracting the non-HTM resistance from the total device resistance, assuming that the resistance of HTM thin films obtained with different spin speeds (1200 rpm and 3000 rpm) is directly proportional to their thickness (around 210 nm for 1200 rpm and 150 nm for 3000 rpm, respectively), and that the non-HTM resistance including Au-HTM contact resistance remains the same. The HTM conductivity was obtained using σ = $d \cdot R^{-1} \cdot A^{-1}$, where *d* is the thickness of the film, and *A* is the effective area (12 mm$^2$). The HTM films were annealed prior to the Au deposition since Au-HTM contact resistance reduces significantly with heating.

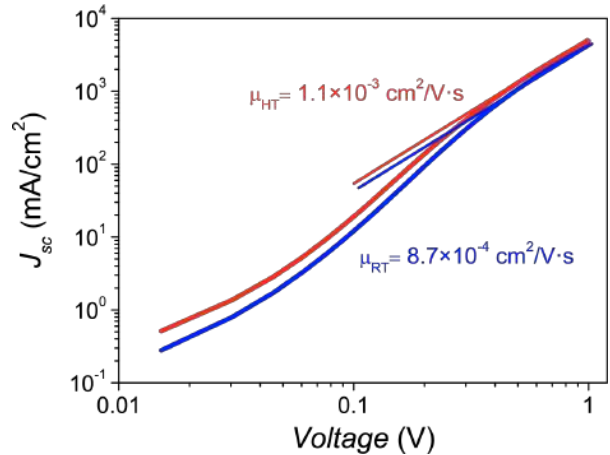

**Fig. S10.** Hole mobility measurements of **DFH** in hole-only devices. The device configuration is ITO/MoO$_3$/HTM/Au, where the HTM layer was either stored at room temperature after spin-coating or annealed at 135 ℃.

**Table S5.** Hole conductivities and mobilities for **DFH** and selected HTMs.

| Condition | $\mu_h$ ($\times 10^{-3}$ cm$^2$·V$^{-1}$·s$^{-1}$)[C] | Conductivity ($\times 10^{-3}$ mS·cm$^{-1}$)[B] |
|---|---|---|
| **DFH** (as-prepared) | 0.9 | 3.5 |
| **DFH** (annealed @ 135 °C) | 1.1 | 7.4 |
| **KR321** | 0.26 | - |
| **PTAA** (doped with Li$^+$) | 0.43 | - |
| **spiro-OMeTAD** (doped with Li$^+$) | 0.69 | 117 |

(A) Hole mobility was measured using the space-charge-limited-current (SCLC) method on hole-only devices (ITO/MoO$_3$/HTM/Au). (B) Conductivity was calculated using σ = $d \cdot R^{-1} \cdot A^{-1}$, where $R$ is the film resistance measured on ITO/HTM/Au, $A$ is the effective area (12 mm$^2$) and, $d$ is the thickness of the HTM thin film. For each condition, two different $d$ each in four devices were measured in order to subtract the contact resistances which are independance to $d$. (C) Thickness ~ 40 nm.

### *Fluorescence Lifetime Measurement*

Experiments were carried out using a customized laser system provided by the LASIR facility at the University of British Columbia. Pulsed (10 Hz) 532 nm laser was used to excite (prompt) perovskite samples on various substrates and the emitted photons were probed by a Hamamatsu dynamic range streak camera (C7700). The device architecture is glass/HTM/perovskite/PMMA.

## Solar Cell Fabrication and Characterization

Methylammonium iodide (MAI) and formamidinium iodide (FAI) were synthesized by the reaction of methylamine and formamidine with hydroiodic acid, respectively, as previously reported. ITO coated glass substrates were obtained from Xin Yan Technology Ltd (2.5 cm × 2.5 cm, $R_s$ = 20 Ω/□). PbI$_2$ (99.9985%), C$_{60}$ (99.5%), bathocuproine (BCP) (98%) and silver (Ag) (99.99%) were purchased from Fisher Scientific, Nano-C, Fisher Scientific, and Kurt J. Lesker, respectively. All materials were used without any further purification.

ITO-coated glass substrates were cleaned by 20 min sonication in each of detergent (extran 300, 2%), deionized water, acetone and isopropanol. After drying in a stream of nitrogen, a 15 min ultraviolet-ozone treatment was carried out immediately prior to the deposition of the hole transport layer (HTL). **DFH** was dissolved in chlorobenzene (10, 15, or 20 mg/mL) and deposited on the ITO substrate by spin-coating (6000 rpm for 30 s). The films were then annealed at various temperatures for 20 min on a hot-plate in ambient air. PTAA HTLs were deposited by spin coating from a 1.5 mg/mL toluene solution, as reported previously[45]. An MAPbI$_3$ perovskite precursor solution (1.2 M) was prepared by mixing MAI and PbI$_2$ (1:1 molar ratio) in anhydrous mixed solvent (4:1 DMF/DMSO). A solution of FAI and PbCl$_2$ was also prepared at the same concentration using the same solvent mixture. The two solutions were combined in a 9:1 ratio (MA:FA, v/v) to produce the final perovskite precursor solution. Perovskite layers were deposited by dropping the precursor solution onto the substrate, and after spinning at 4000 rpm for 7 s, dropping anhydrous chlorobenzene (120 μL) onto the center of the substrate. The substrate was spun for a further 30 s at 4000 rpm without pause. The transparent yellow films obtained after spin coating were heated at 35 °C for 20 min, after which they became black. The films were further annealed at 85 °C for 10 min. To complete the device stack, C$_{60}$ (40 nm), BCP (8 nm) and Ag (100 nm) were sequentially deposited by thermal evaporation at a base pressure of $1 \times 10^{-6}$ mbar.

Current-voltage curves of perovskite solar cells and hole-only devices were recorded with a Keithley 2400 source-measure unit. Devices were measured in the glovebox (H$_2$O < 1 ppm, O$_2$ < 0.1 ppm) using a 450 W Class AAA solar simulator equipped with a AM1.5G filter (Sol3A, Oriel Instruments). Before measuring, a standard silicon reference cell (91150V, Oriel Instruments) was used to set the light intensity to 1 sun. During the measurement, the cell was covered by a non-reflective metal mask with an aperture of 0.0708 cm$^2$. Incident photon to current (IPCE) measurements were performed in air using a QE-PV-SI system (Oriel Instruments) consisting of a 300 W Xe arc lamp, monochromator, chopper, lock-in amplifier and certified silicon reference cell, operating at a 30 Hz beam-chopping frequency. Scanning Electron Microscopy (SEM) images were acquired on either a FEI Helios NanoLab 650 dual beam SEM at 5 kV and 50 pA or a Hitachi SU8010 microscope at 3 kV and 10 pA using a through-lens detector in secondary electron mode.

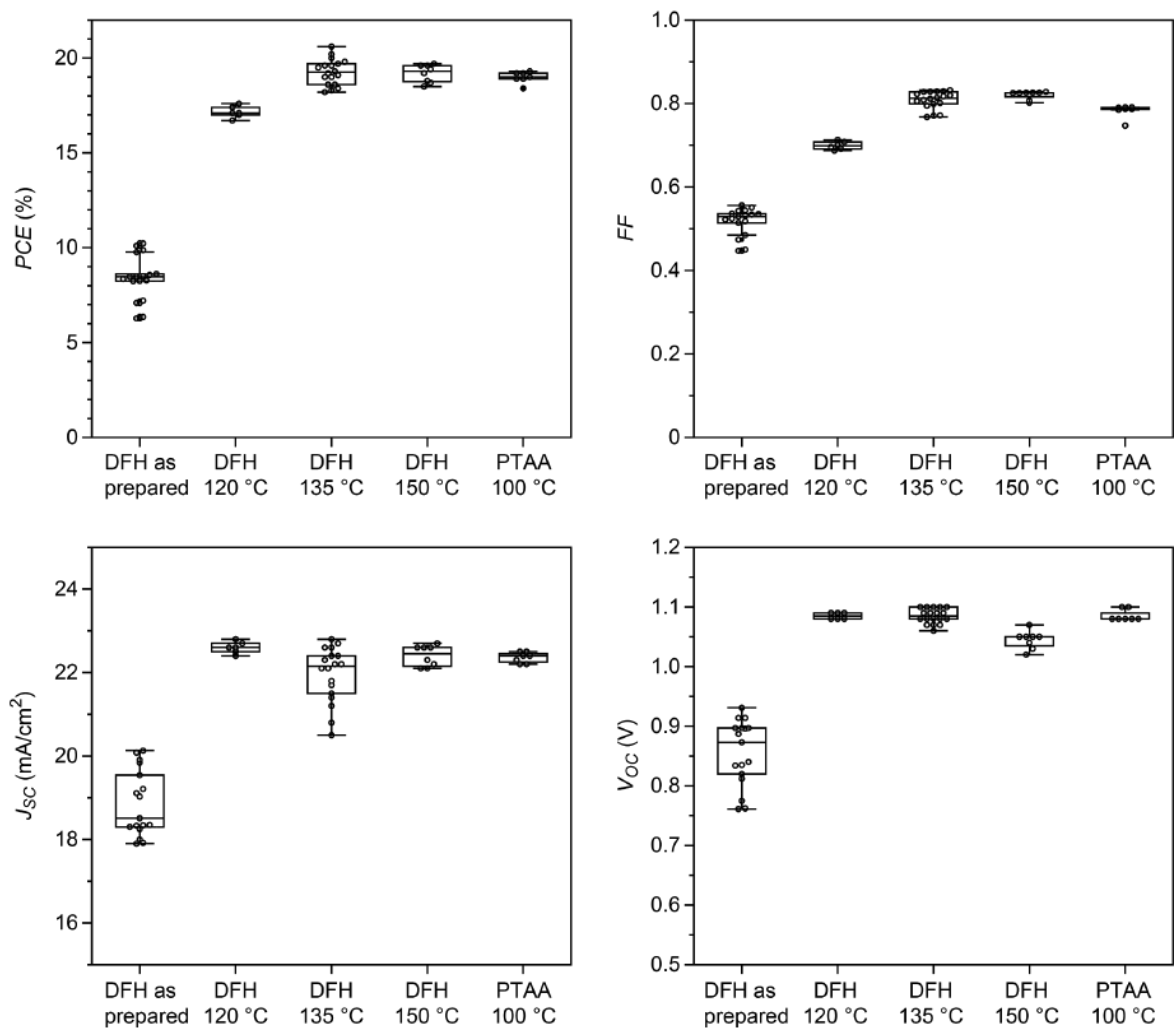

**Fig. S11**. Device performance statistics for PSCs with different hole transport layers obtained by reverse $J-V$ scans. Sample sizes are 17, 6, 18, 8 and 7 for devices with as prepared DFH, 120 °C annealed DFH, 135 °C annealed DFH, 150 °C annealed DFH and 100 °C annealed PTAA as hole transport layers. PTAA was annealed at 100 °C according to device optimization in literature.[7]

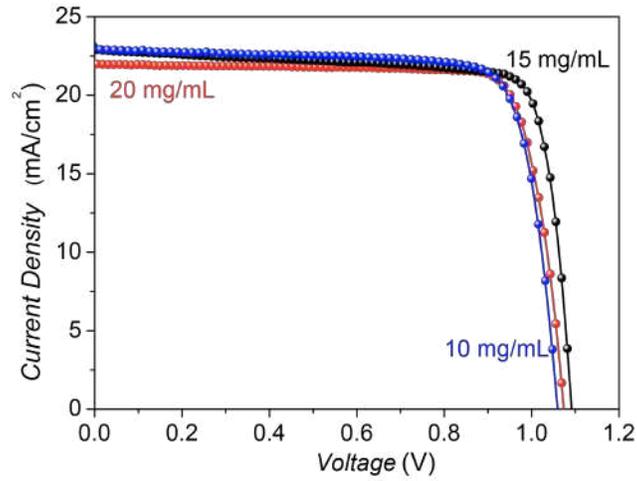

**Fig. S12.** *J-V* curves of PSCs with varying thickness of **DFH** layers, illustrating the effect of **DFH** thickness on PSC performance.

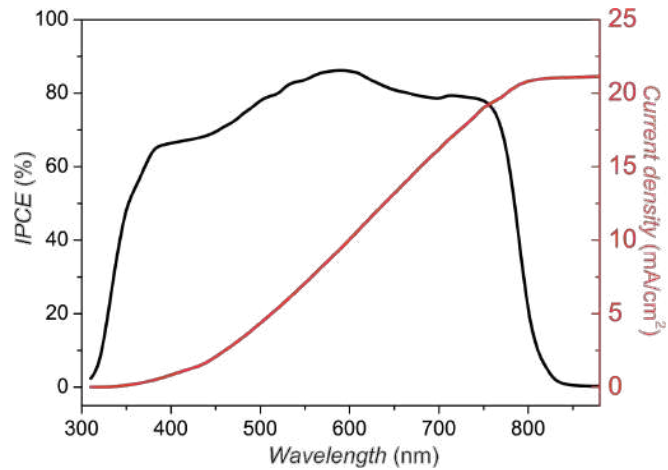

**Fig. S13.** IPCE spectrum of a PSC with a **DFH** HTL annealed at 135 °C.

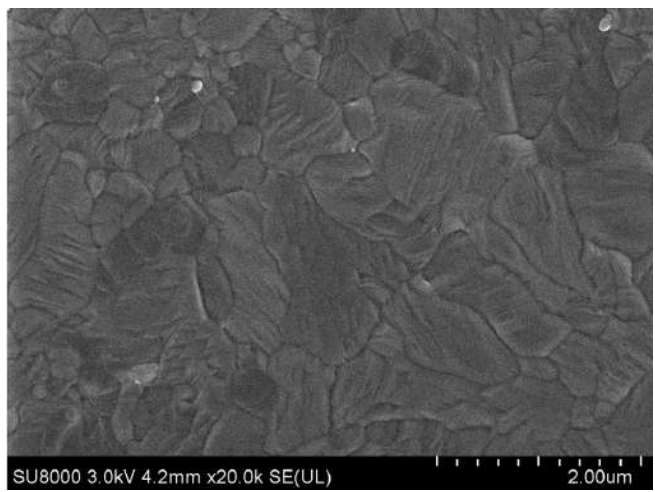

**Fig. S14.** SEM images of perovskite layers grown on a thin film of 135 °C annealed **DFH**. The scale bar is 2 μm.

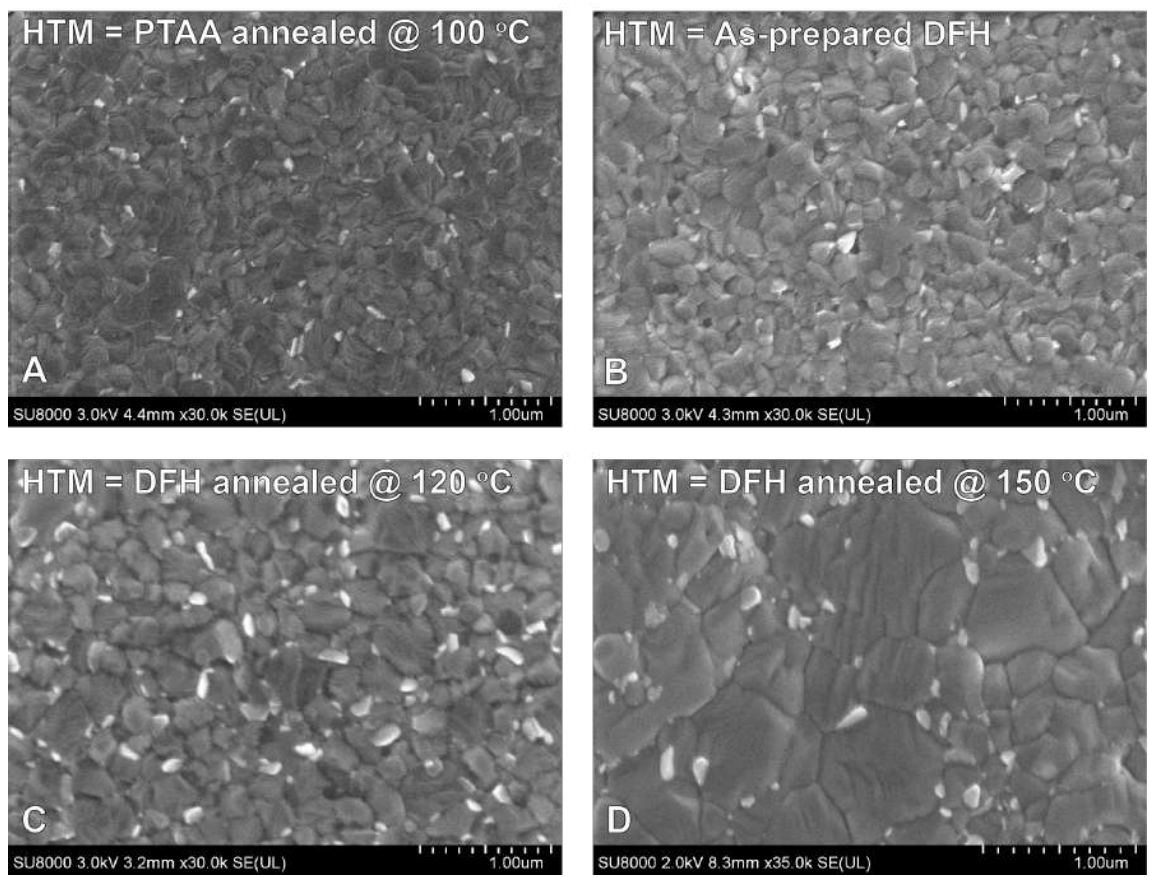

**Fig. S15.** SEM images of perovskite layers grown on thin films of (A) 100 °C annealed PTAA (B) as-prepared and unheated DFH and (C) 120 °C annealed **DFH** and (D) 135 °C annealed **DFH**. The scale bars are 1 μm.

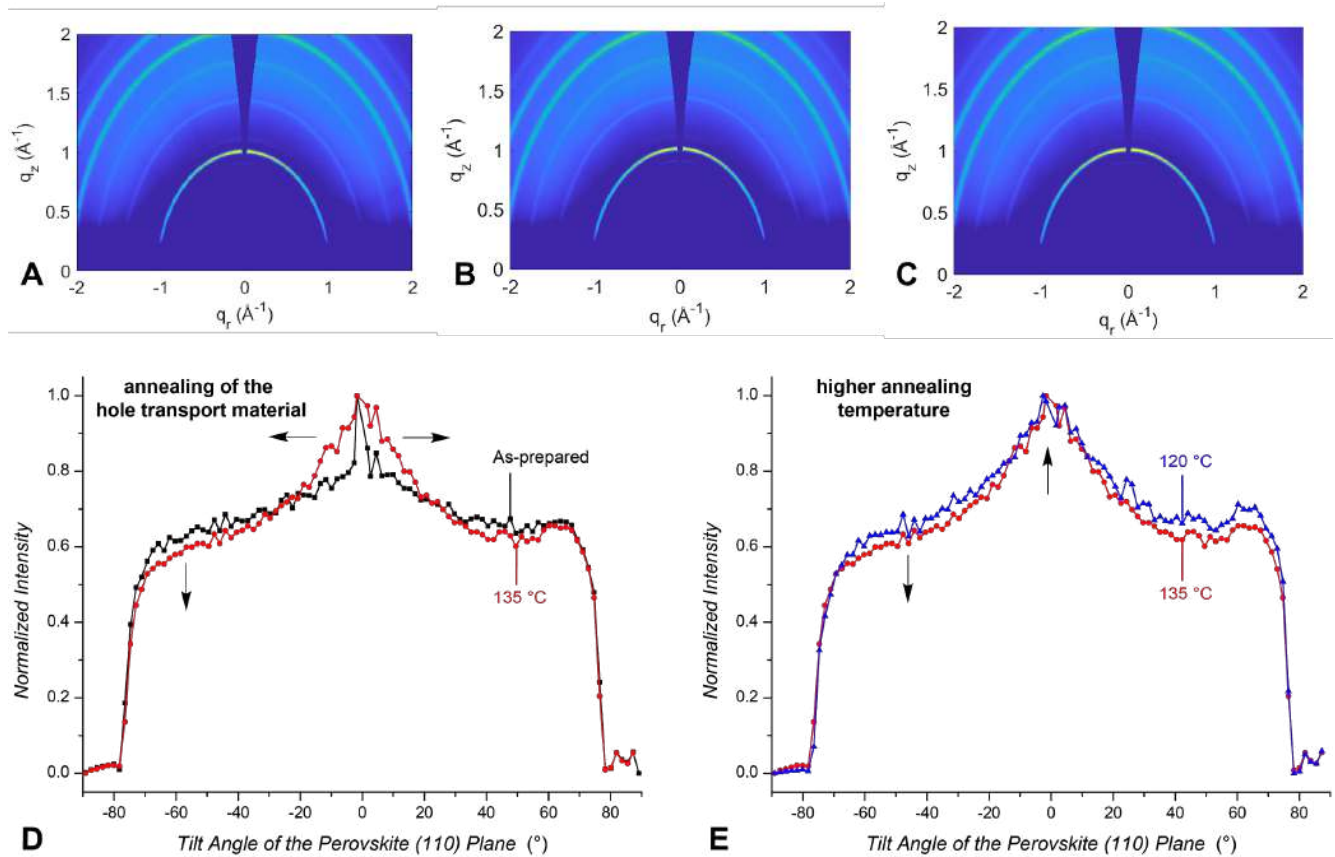

**Fig. S16**. GIXD patterns of perovskite layers grown on **DFH** thin films (A) as-prepared (B) annealed at 120 ℃, and (C) annealed at 135 °C. Radially integrated intensity plots of the GIXRD patterns along the $q = 1.0106 \text{ Å}^{-1}$ ring assigned to the (110) plane showing the differences between the perovskite layers grown on (D) as-prepared and 135 °C annealed **DFH**, and (E) 120 °C and 135 °C annealed **DFH**.

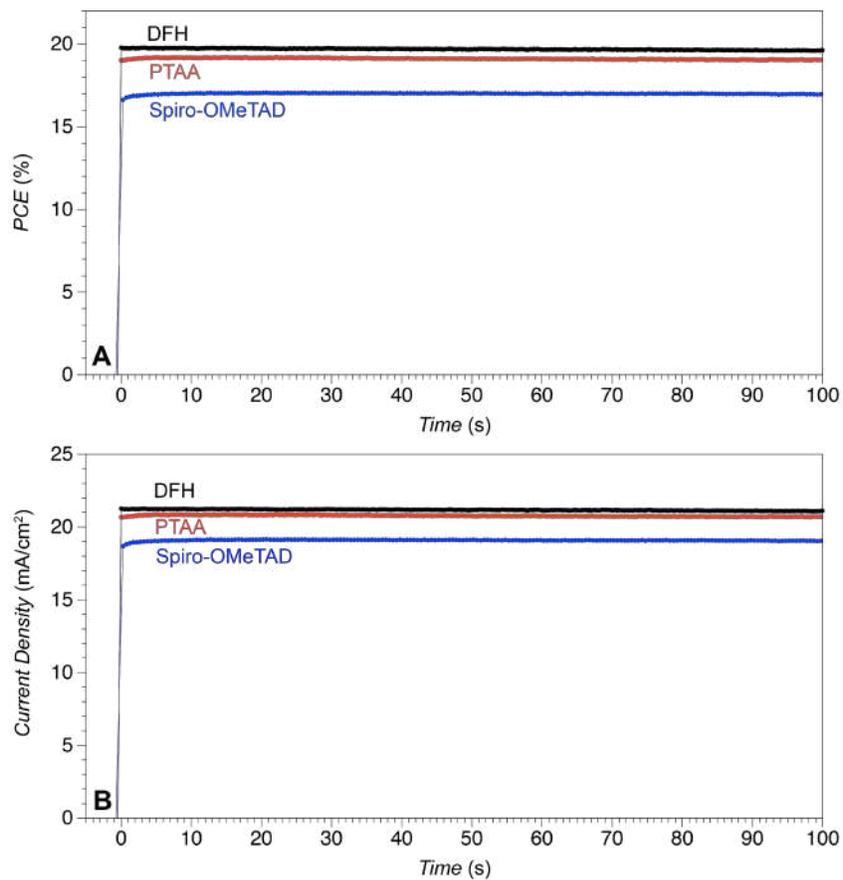

**Fig. S17.** Stabilized (A) *PCE* and (B) current density of PSCs with 135°C annealed hole transport layers measured at the maximum power points every 0.25 seconds, the potentials for DFH, PTAA and Spiro-OMeTAD are 0.93 V, 0.92 V and 0.89 V, respectively.

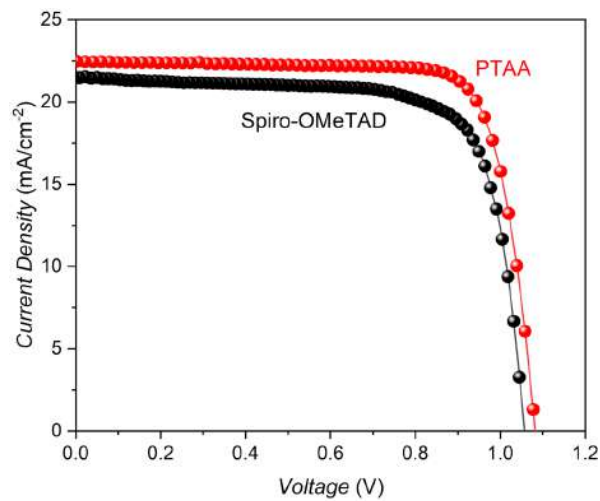

**Fig.S18.** Reverse J−V scans of champion PSCs with PTAA and Spiro-OMeTAD hole transport layers.

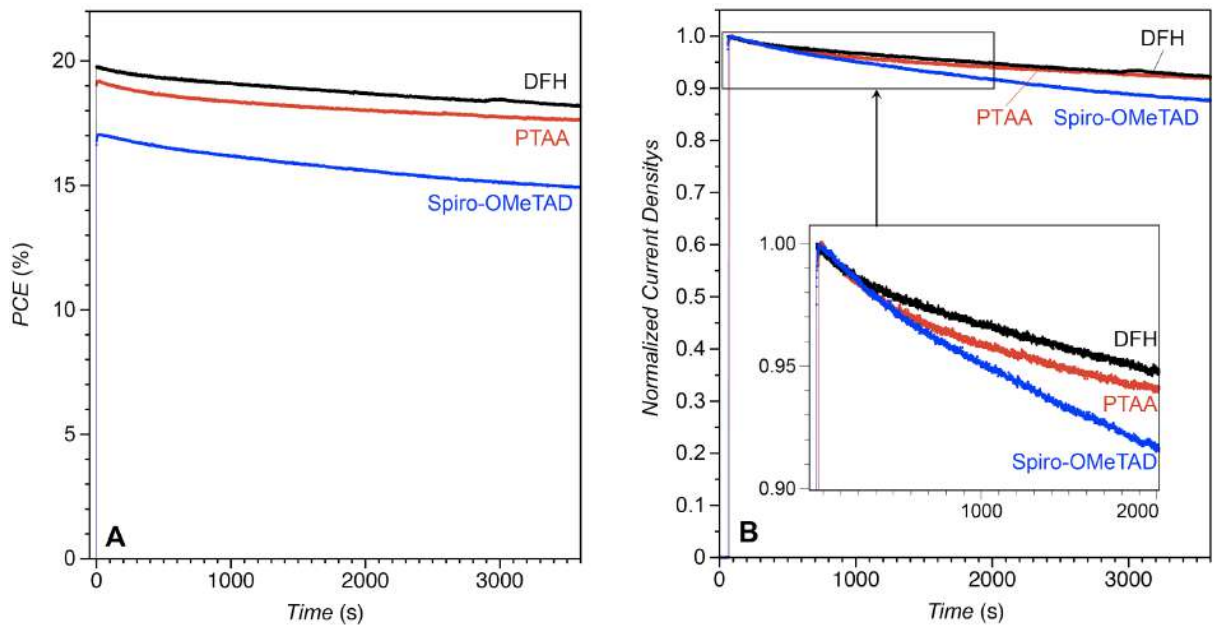

**Fig. S19.** Operational stability in terms of the decay of (A) *PCE* and (B) normalized current density of perovskite solar cells under 1-sun condition loaded with a constant voltage of their initial maximum power points (0.93 V, 0.92 V and 0.89 V for **DFH**, PTAA and Spiro-OMeTAD, respectively).

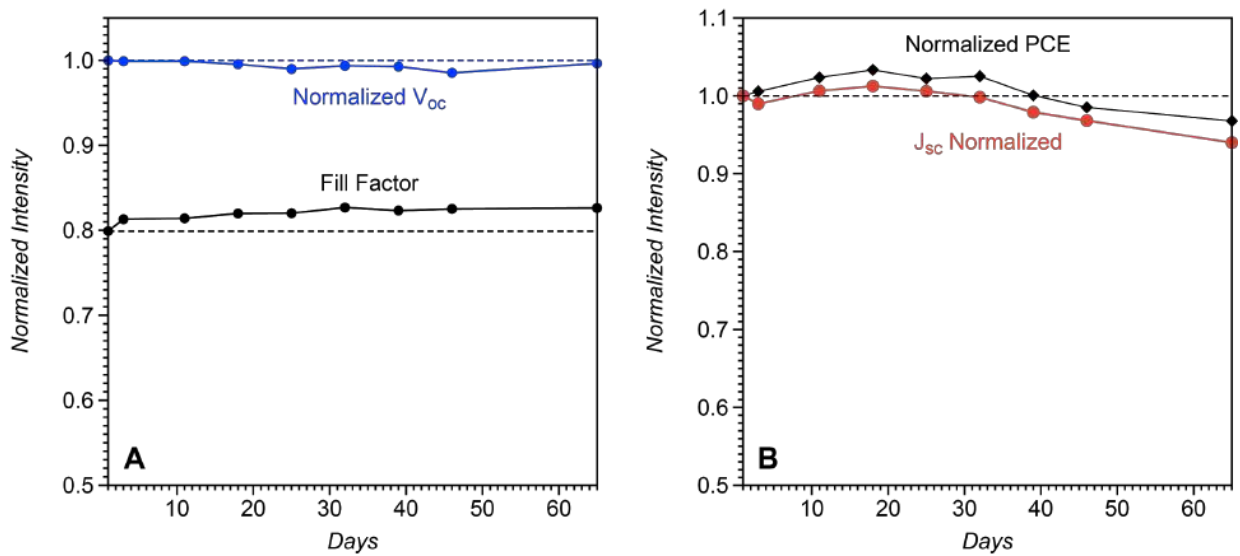

**Fig. S20.** Storage stability of a PSC device with 135 °C annealed **DFH**, under an inert atmosphere. Dashed lines indicate data collected on day 1. Initial *PCE* = 19.3%, $V_{oc}$ = 1.09 V; *FF* = 0.79.

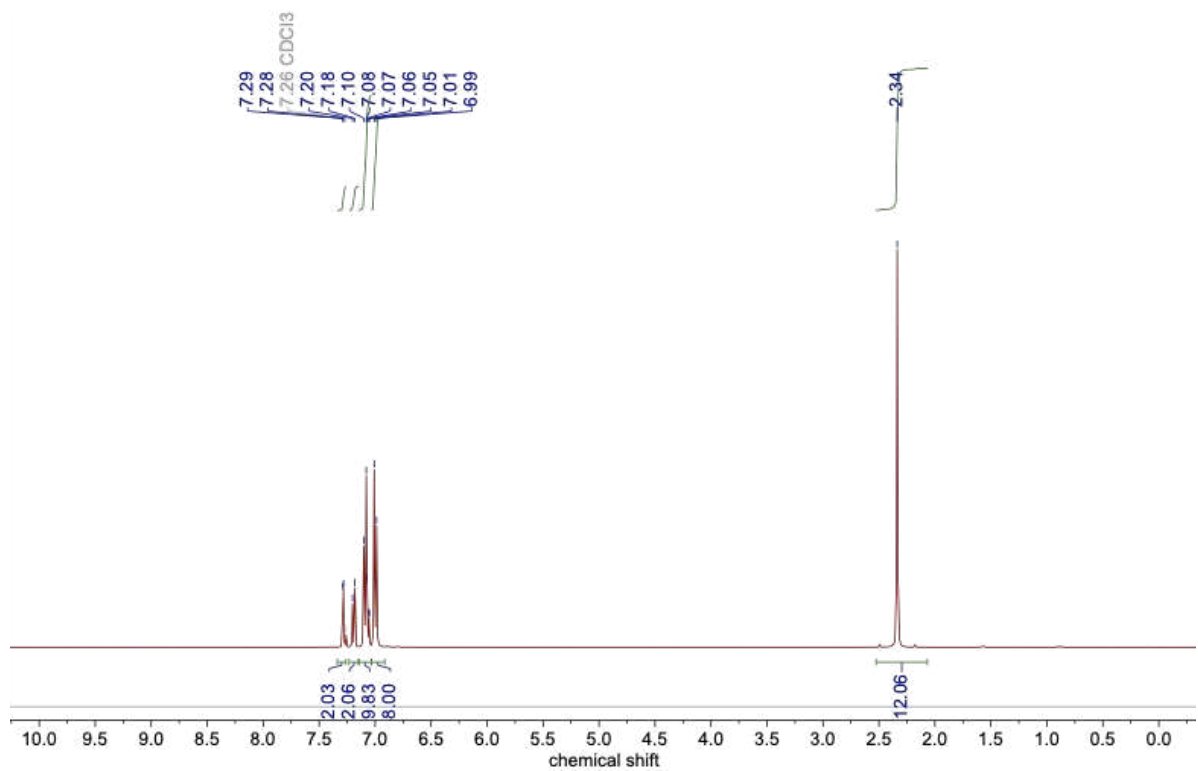

**Fig. NMR1.** $^{1}$H NMR spectrum of **P1** in CDCl$_3$.

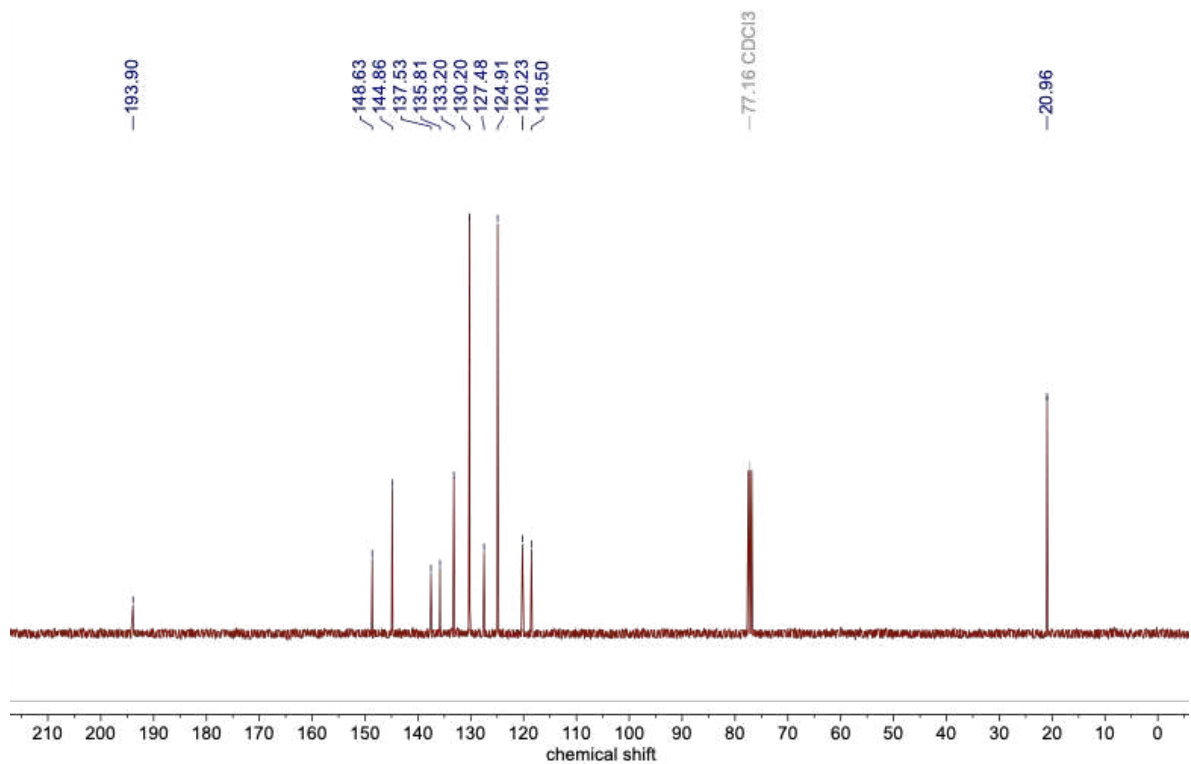

**Fig. NMR2.** $^{13}$C{$^{1}$H} NMR spectrum of **P1** in CDCl$_3$

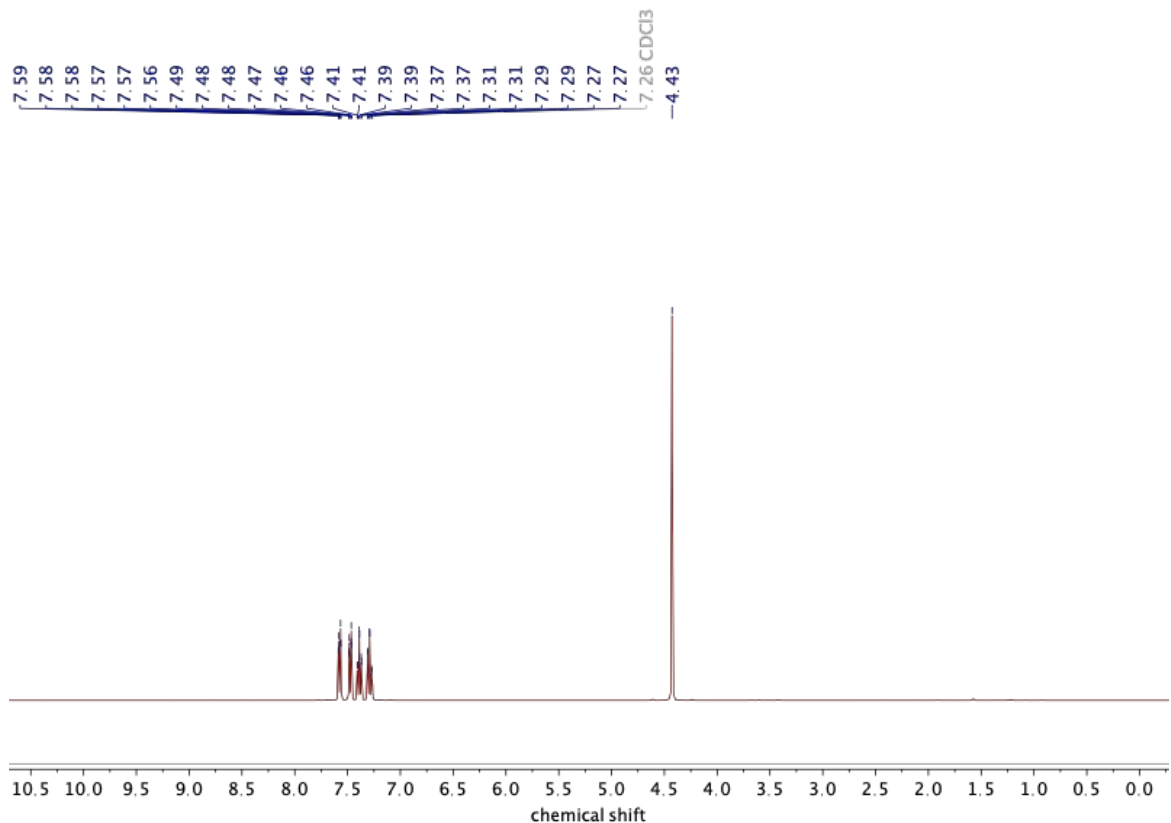

**Fig. NMR3.** ¹H NMR spectrum of **P2** in CD₂Cl₂.

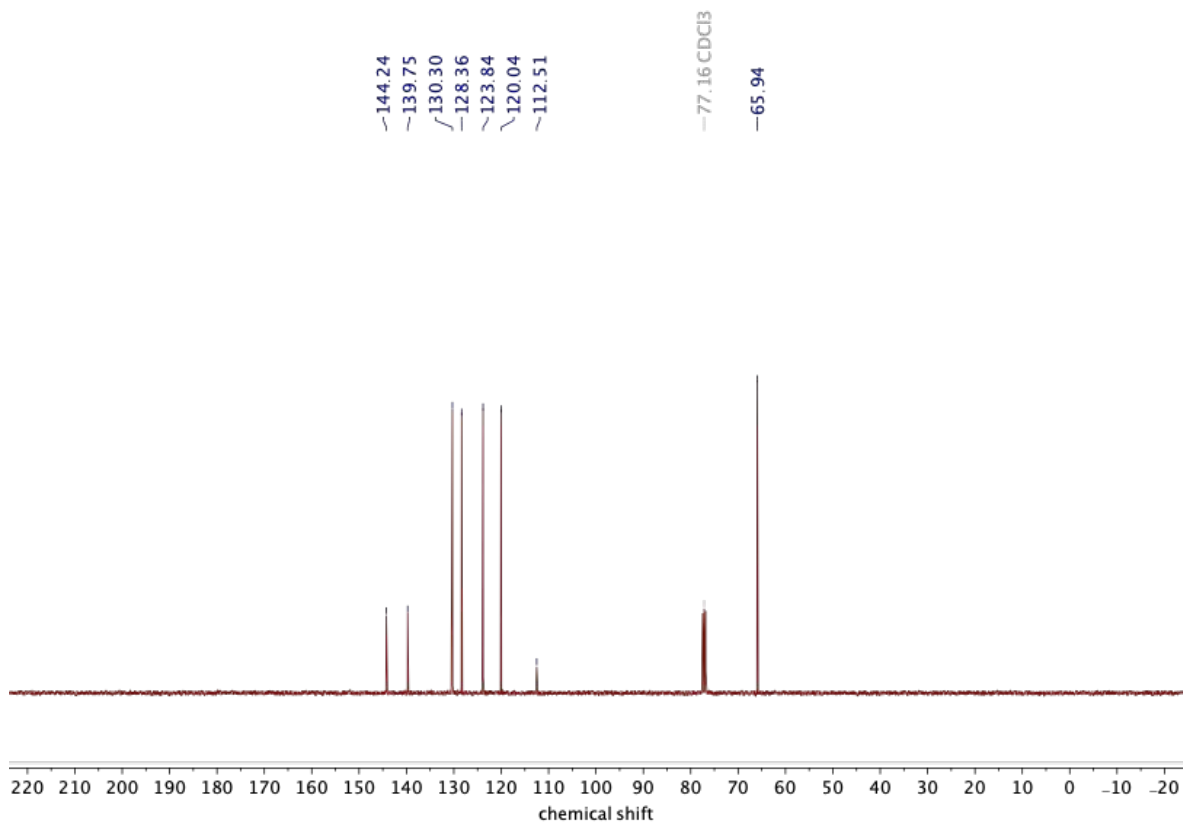

**Fig. NMR4.** ¹³C{¹H} NMR spectrum of **P2** in CD₂Cl₂.

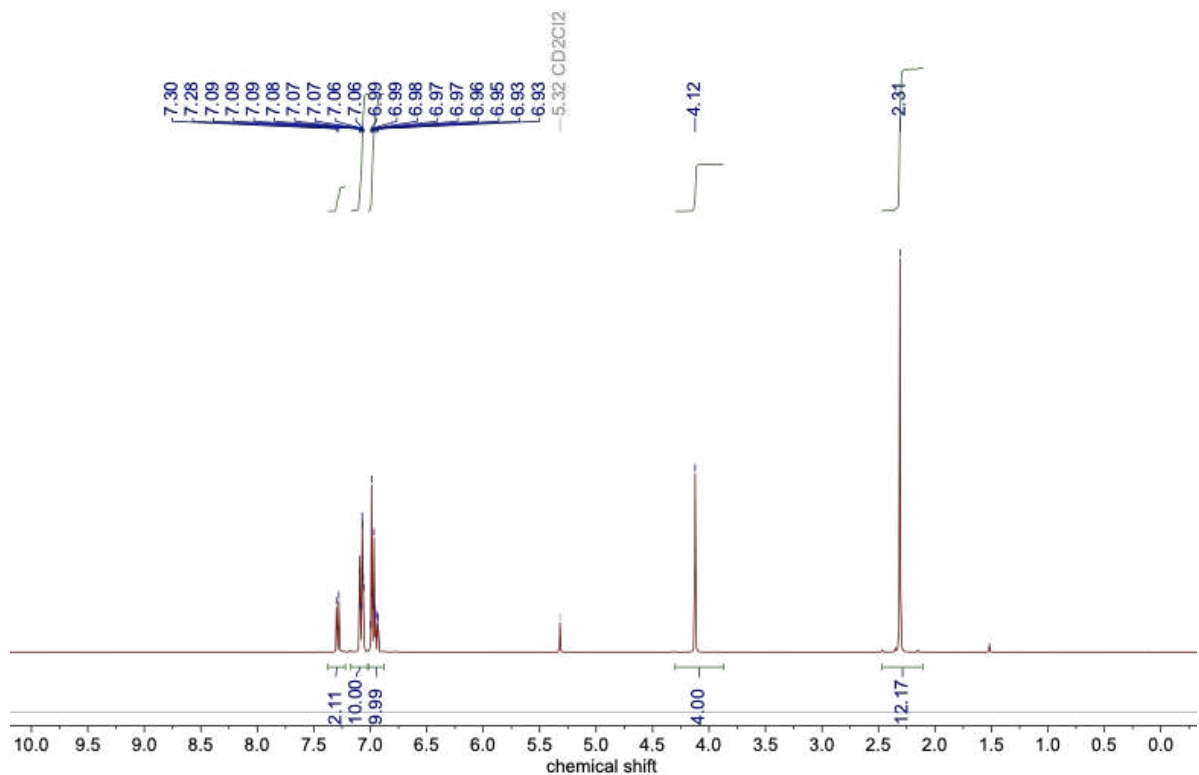

**Fig. NMR5.** $^1$H NMR spectrum of **DFH** in CD$_2$Cl$_2$.

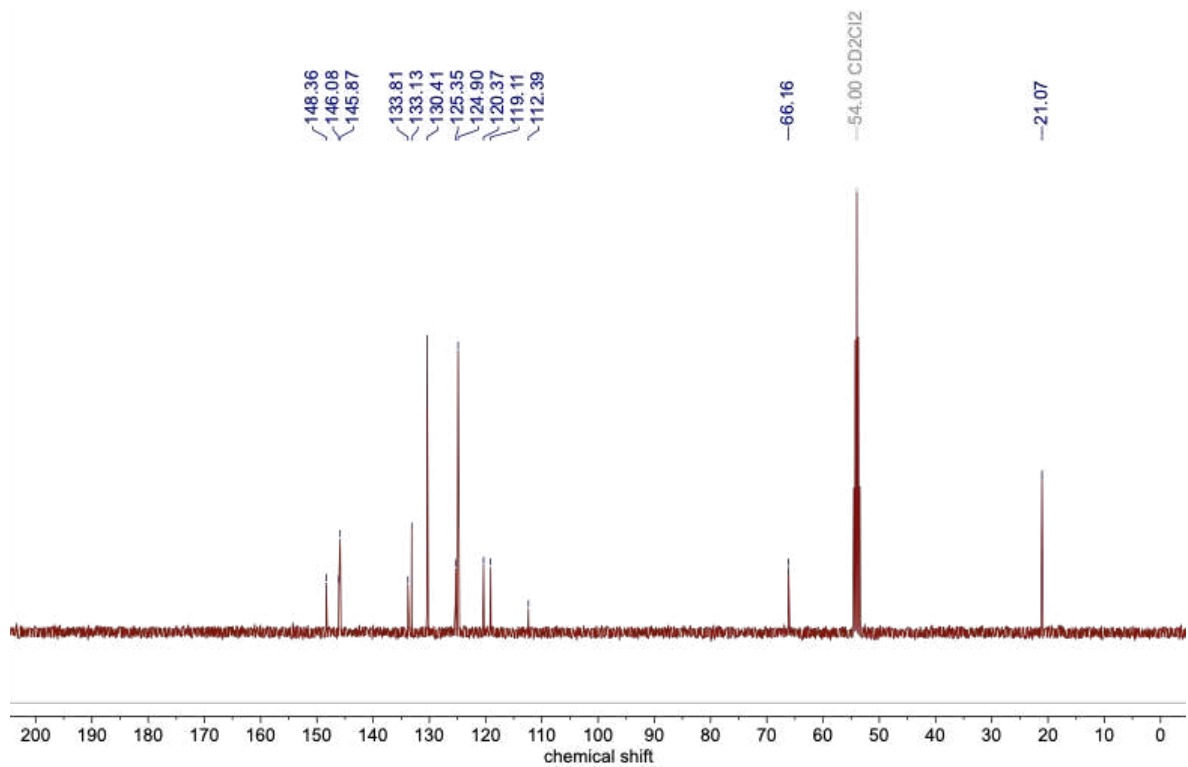

**Fig. NMR6.** $^{13}$C{$^1$H} NMR spectrum of **DFH** in CD$_2$Cl$_2$.

**References :**

1   G. M. Sheldrick, *Acta Crystallogr. A*, 2015, **71**, 3–8.

2   O. V. Dolomanov, L. J. Bourhis, R. J. Gildea, J. A. K. Howard and H. Puschmann, *J. Appl. Crystallogr.*, 2009, **42**, 339–341.

3   M. J. Frisch, G. W. Trucks, H. B. Schlegel, G. E. Scuseria, M. A. Robb, J. R. Cheeseman, G. Scalmani, V. Barone, G. A. Petersson, H. Nakatsuji, X. Li, M. Caricato, A. Marenich, J. Bloino, B. G. Janesko, R. Gomperts, B. Mennucci, H. P. Hratchian, J. V. Ortiz, A. F. Izmaylov, J. L. Sonnenberg, D. Williams-Young, F. Ding, F. Lipparini, F. Egidi, J. Goings, B. Peng, A. Petrone, T. Henderson, D. Ranasinghe, V. G. Zakrzewski, J. Gao, N. Rega, G. Zheng, W. Liang, M. Hada, M. Ehara, K. Toyota, R. Fukuda, J. Hasegawa, M. Ishida, T. Nakajima, Y. Honda, O. Kitao, H. Nakai, T. Vreven, K. Throssell, J. A. Montgomery, J. E. Peralta, F. Ogliaro, M. Bearpark, J. J. Heyd, E. Brothers, K. N. Kudin, V. N. Staroverov, T. Keith, R. Kobayashi, J. Normand, K. Raghavachari, A. Rendell, J. C. Burant, S. S. Iyengar, J. Tomasi, M. Cossi, J. M. Millam, M. Klene, C. Adamo, R. Cammi, J. W. Ochterski, R. L. Martin, K. Morokuma, O. Farkas, J. B. Foresman, and D. J. Fox, Gaussian, Inc. , Wallingford CT, 2016., *Gaussian 09 Rev-D.01*, .

4   J.-D. Chai and M. Head-Gordon, *Phys. Chem. Chem. Phys.*, 2008, **10**, 6615–6620.

5   Y. Li, H. Li, C. Zhong, G. Sini and J.-L. Brédas, *npj Flexible Electronics*, 2017, **1**, 2.

6   J. B. Foresman and Æ. Frisch., *Exploring Chemistry with Electronic Structure Methods*, 3rd ed. (Gaussian, Inc., Wallingford, CT, 2015).

7   M. Stolterfoht, C. M. Wolff, Y. Amir, A. Paulke, L. Perdigón-Toro, P. Caprioglio and D. Neher, *Energy Environ. Sci.*, 2017, **10**, 1530–1539.